\documentclass[12pt]{article}
\pdfoutput=1
\usepackage{tikz}
\usetikzlibrary{matrix,arrows,positioning,shapes,snakes,fadings,decorations.pathmorphing,
decorations.pathreplacing,automata,shadows,decorations.markings,patterns}
\usepackage[thicklines]{cancel}
\usetikzlibrary{calc,shapes.callouts,shapes.arrows}
\usepackage{mathtools}
\usepackage{mathpazo}
\usepackage{jheppub}
\usepackage{amsmath,amssymb,euscript,array,mathrsfs,appendix,ctable,marvosym}
\usepackage{arydshln}
\usepackage{todonotes}
\usepackage{graphicx}
\usepackage[normalem]{ulem}

\newcommand\blank[1]{#1}
\renewcommand\blank[1]{}
\def\Buildrel#1\over#2\under#3{\mathrel{\mathop{\kern0pt
#2}\limits^{#1}_{#3}}}

\newcommand*\circled[1]{\footnotesize\tikz[baseline=(char.base)]{%
            \node[shape=circle,fill=black!20,draw,inner sep=2pt] (char) {#1};}}

\usepackage{enumitem}

\def\Lax{{\mathscr L}}
\def\CF{{\cal F}}

\def\JJ{\mathscr{J}}

\def\Pr{{\mathbb P}}

\def\msu{\mathfrak{su}}

\def\p{\pi}

\newcommand{\sech}{\operatorname{sech}}

\newcommand{\Tr}{\operatorname{Tr}}

\newcommand{\IM}{\operatorname{Im}}

\def\B0{{\boldsymbol 0}}

\def\BId{{\boldsymbol{I}}}
\def\BR{{\boldsymbol{\cal R}}}
\def\BOmega{{\boldsymbol\Omega}}

\def\BTheta{{\boldsymbol\Theta}}

\def\SU{\text{SU}}
\def\U{\text{U}}

\def\ee{\boldsymbol{e}}

\def\Dbarslash{\,\,{\raise.15ex\hbox{/}\mkern-12mu {\bar D}}}
\def\Dslash{\,\,{\raise.15ex\hbox{/}\mkern-12mu D}}
\def\delslash{\,\,{\raise.15ex\hbox{/}\mkern-9mu \partial}}
\def\delbarslash{\,\,{\raise.15ex\hbox{/}\mkern-9mu {\bar\partial}}}

\def\ket#1{| #1\rangle}

\newcommand{\MAT}[1]{\begin{pmatrix} #1\end{pmatrix}}

\newcommand{\EQ}[1]{\begin{equation}\begin{split} #1
\end{split}\end{equation}}

\newcommand{\FIG}[1]{\begin{figure}[ht]\begin{center} #1 \end{center}\end{figure}}

\def\be{\begin{equation}}
\def\ee{\end{equation}}
\newcommand{\DT}[1]{ {\textcolor{cyan}{#1}}}
\newcommand \arxivlink [1]{\href{http://arxiv.org/abs/#1}{\tt arXiv:#1}}
\newcommand{\WW}[5] {W\left.\left.\left(\hspace{-0.2cm}\raisebox{-14.5pt}{\begin{tikzpicture}[scale=0.32]
\node at (0,1.1) {$#4$};
\node at (0,-1.1) {$#2$};
\node at (1.2,0) {$#3$};
\node at (-1.2,0) {$#1$};
\end{tikzpicture}}\hspace{-0.15cm}\right.\right| #5\right)}

\title{Quantum Anisotropic Sigma and Lambda Models as Spin Chains}
\author{Calan Appadu, Timothy J. Hollowood, Dafydd Price and Daniel C. Thompson}
\affiliation{Department of Physics, Swansea University, Swansea, SA2 8PP, U.K.}

\emailAdd{t.hollowood@swansea.ac.uk}
\emailAdd{D.C.Thompson@Swansea.ac.uk}

\abstract{We consider lambda and anisotropic deformations of the $\SU(2)$ principal chiral model and show how they can be quantized in the Hamiltonian formalism on a lattice as a suitable spin chain. The spin chain is related to the higher spin XXZ Heisenberg chain and can be solved by using the Bethe Ansatz. This yields the spectrum and S-matrix of the excitations. In particular, we find the S-matrix in the gapped anti-ferromagnetic regime. In this regime, a continuum limit does not exist and this suggests that the field theories in this regime, precisely ones with a cyclic RG like the Yang-Baxter deformations, may only exist as effective theories. In a certain limit, we show that the XXZ type lambda model gives the symmetric space $\SU(2)/\U(1)$ lambda model and, hence, we are able to find its spectrum and S-matrix and show that it gives the S-matrix of the $\text{O}(3)$ sigma model in the appropriate limit. Finally, we show the full consistency of the S-matrix and the Lagrangian formulations of the lambda model, by coupling to a conserved charge and computing the way the ground state energy changes in both pictures.}

\notoc
\begin{document}

\maketitle

\newpage

\tableofcontents
\section{Introduction}

Integrable QFTs in $1+1$ dimension play an important r\^ole in diverse areas of physics from condensed matter to string theory.  Notably in the gauge/gravity correspondence  the world sheet theory describing strings in an $AdS_5 \times S^5$ background is a type of integrable sigma model \cite{Bena:2003wd}.  This in turn has prompted tremendous progress in using integrability to elucidate the properties of the corresponding dual gauge theory (for a review \cite{Beisert:2010jr}).   This success has motivated substantial activity in exploring deformations of (string) sigma models which preserve integrability.  Our present focus will be on a class of integrable theories developed in \cite{Sfetsos:2013wia,Hollowood:2014rla,Hollowood:2014qma} known as {\it lambda models\/} that can be associated to an integrable sigma model.

There is by now significant evidence that lambda models give deformations that are consistent string theories \cite{Sfetsos:2014cea,Demulder:2015lva,Appadu:2015nfa}.  Based on the displayed symmetries, and the expectation that  classical integrability holds at the quantum level, exact S-matrices obeying the axioms of factorised scattering have been proposed for  lambda models \cite{Hollowood:2015dpa}.  The idea of lambda model deformation of an integrable sigma model goes back a long way to various attempts to solve the integrable sigma models by Bethe Ansatz techniques \cite{Polyakov:1983tt,Faddeev:1985qu,Evans:1994hi}.

Whilst the use of  Bethe Ansatz techniques is certainly compelling it masks one important and outstanding challenge  --  to derive from the classical integrable structure the quantum S-matrix.  In other words, one should like to directly quantize the string world sheet in the Hamiltonian/Hilbert space formalism via the Quantum Inverse Scattering Method (QISM).    One significant block to applying this method is that the classical theory is {\em non ultra-local} i.e. the classical Poisson brackets yield terms with derivatives of delta functions.   As we shall describe in more detail in section \ref{sec:nonultra} this prevents a straightforward quantisation.   Ultimately the goal should be to resolve this for the $AdS_5 \times S^5$ string world sheet theory, building on the {\it alleviation procedure\/} of Delduc et al.~\cite{Delduc:2012vq}. Here we shall restrict our attention to the simpler setting of bosonic sigma models where there is already   much to be learnt.

 It turns out, at least for bosonic sigma and lambda models, that the non ultra-locality problem can be overcome \cite{Appadu:2017fff} and the quantisation achieved.  In \cite{Appadu:2017fff}   the   lambda model corresponding to an integrable bosonic Principal Chiral Model for the group $\SU(N)$ was  quantised  whilst maintaining integrability using  a procedure of  Faddeev-Reshitikhin   \cite{Faddeev:1985qu}.  The theory regulated on a light-cone lattice  is then described by an integrable quantum spin chain.  For the case of $\SU(2)$ this is nothing more than a higher spin $XXX_\frac{k}{2}$ version of the familiar Heisenberg $XXX_\frac{1}{2}$  spin chain \cite{Heisenberg:1928,Dirac:1929} whose solution  goes back to Bethe \cite{Bethe:1931hc}.
  
There is a well known generalisation of this spin chain to the non-isotropic $XXZ_\frac{k}{2}$ spin chain which can either be in  a  ferromagentic phase, a gapped anti-ferromagentic phase or a gapless paramagnetic phase depending on the anisotropy parameter (and the applied magnetic field).   It is natural   to ask is if this then provides a quantisation for an integrable theory akin to the lambda model?  A key aim in this work is to answer in the affirmative by showing that the $XXZ_\frac{k}{2}$ spin chain provides a quantisation for a two-parameter integrable model known -- with no coincidence -- as the XXZ-lambda model.\footnote{There is a further generalisation of spin chains to the $XYX$ type  which one anticipates would correspond to the $XYZ$ lambda model with an elliptic Lax connection  --  we intend to say more of in a future note.} 
 
The XXZ-lambda model can be thought of as the integrable lambda model associated to the integrable sigma model with target space the squashed three-sphere.  In a recent work \cite{Appadu:2017bnv} we examined this model which is described by two   RG variant parameters, $\lambda$ and $\xi$, and an integer quantised level $k$.  Both the symmetry structure of the theory and its behaviour   under renormalisation depend crucially on whether $\lambda < \xi$ or not.   In the  $\lambda>\xi$ regime the theory is defined by a UV CFT fixed point but in the $\lambda<\xi$ regime the theory, if it were to be well defined at all scales, appears to display a surprising cyclic RG behaviour.   From the symmetries of the  XXZ-lambda model in \cite{Appadu:2017bnv} we were able to conjecture exact S-matrices for both these regimes.

In this paper we aim to provide a precise study of the quantisation of the XXZ-lambda model, test the conjectured S-matrices of \cite{Appadu:2017bnv} and resolve the physical understanding of behaviour of the model in its different regimes. More precisely in this paper:
\begin{enumerate}[label=\protect\circled{\arabic*}]
\item We shall apply the Faddeev-Reshitikhin procedure \cite{Faddeev:1985qu} to identify a spin chain associated to XXZ lambda model. This spin chain is a higher spin $s=k/2$ Heisenberg XXZ type spin chain.
\item We show that the spectrum and S-matrix of the lowest lying excitations of the XXZ spin chain in the paramagnetic regime matches that conjectured in the XXZ lambda model in the regime $\lambda>\xi$ based on the symmetries.
\item We further provide a direct connection between the Lagrangian formulation of the XXZ lambda model, the S-matrix and the spin chain by calculating the response of the free energy to a chemical potential for a particular $\U(1)$ symmetry. This involve a chemical potential, or finite density, type Thermal Bethe Ansatz (TBA) calculation. 
\item We show that the limit in which the XXZ lambda model  has an emergent $\U(1)$ gauge symmetry and reduces to the $\SU(2)/\U(1)$ lambda model is reflected in the S-matrix through the decoupling of solitons and heavy bound states leaving just an $\text{O}(3)$ triplet's worth of the lightest breather modes as physical states.
\item We also describe the $\lambda<\xi$ regime of the lambda model which we show is related to the higher spin XXZ spin chain in the anti-ferromagnetic regime. The spin chain in this regime has a gap. Based on a TBA calculation we identify the S-matrix of the lowest lying excitations. This S-matrix involves a new kind of elliptic RSOS piece that we explain in detail. The S-matrix exhibits a periodicity in the rapidity difference of the incoming states. 
\item We then show that this S-matrix coming from the spin chain is subtly different to that expected of the   $\lambda<\xi$ regime  of the XXZ lambda theory.   The latter has a cyclic RG flow and based on the 
difference with the spin chain S-matrix, we argue that it is likely that in the 
$\lambda<\xi$ regime, the XXZ lambda and its sigma model limit, and hence the $\SU(2)$ Yang-Baxter deformed PCM, does not have a continuum limit that retains integrability. \end{enumerate}

The structure of the paper is as follows: In section  \ref{sec:nonultra} we provide an introduction to the problem of quantisation of non ultra-local theories and the overall approach to alleviating non-ultra locality employed in this paper.  In section \ref{sec:genint} we provide a telegraphic review of the integrable field theories that are studied in this paper.   
Section  \ref{sec:FR} shows how the Faddeev-Reshitikhin procedure is applied to the XXZ lambda model.  Section \ref{sec:spinchain} studies the resultant spin chain in some detail allowing us in section \ref{sec:S-matrices} to determine the S-matrices of the XXZ lambda model in its different regimes.  In section \ref{s6} we consider a particular limit where the XXZ lambda model  has an emergent U(1) gauge symmetry  and reduces to the $\SU(2)/\U(1)$ lambda model.  This is reflected in the decoupling of solitons from the spectrum leaving just the lightest breather modes.  In section \ref{sec:TBA} we test the S-matrices determined by using  finite density thermodynamic Bethe Ansatz techniques to  calculate the ground state energy which can then be compared with conventional perturbation theory. 
  
\section{Quantisation, QISM and non ultra-locality}
\label{sec:nonultra}

The quantization of classical relativistic field theories is a very subtle problem. It is not simply a problem of replacing Poisson brackets by commutators of operators because Quantum Field Theories (QFT) have to regularized in order to isolate UV divergences and then the continuum theory defined by a limiting procedure. From the outset, it may not be clear whether the limit actually exists. For QFTs that are used in particle physics, it is important to recognise that it is not necessary to actually take the limit, since these QFT can just be viewed as phenomenological theories that are limited to describe phenomena on length scales much greater than the cut off. 

After the work of Wilson  \cite{Wilson:1974mb}, it now understood that in order to take a continuum limit, one needs a fixed point of the renormalization group (RG). The continuum limit involves defining a series of regularized QFTs with finer and finer cut offs which asymptote to the basin of attraction of the fixed point under renormalization group flow. These basins have infinite dimension and so there is considerable freedom in defining the regularized theory. It is for this reason that it unnecessary and possibly futile to try to actually lift the Poisson brackets of the classical field theory to commutation relations of the regularized theory before taking the continuum limit. Theories which lie in the same basin of attraction are in the same {\it universality class\/} and all give the same QFT in the continuum limit.

Wilson's RG programme is usually set up in the Euclidean QFT where all the symmetries are manifest. However, this begs the question of how renormalization plays out in a Hamiltonian, Hilbert space, formulation of the theory? It is extremely difficult to push the formalism through all the way to arrive at a Hilbert space picture of a non-trivial (i.e.~fully interacting) QFT.

Integrable QFTs in $1+1$ dimensions offer a playground that includes the Hilbert space formalism of interacting QFTs. Integrability is special, of course, but the theories are still far from trivial. The theories we have in mind are Wess-Zumino-Witten (WZW) QFTs  \cite{Witten:1983ar} and their integrable deformations. In this work we shall focus on those associated to the $\SU(2)$ group.  At the classical level, the currents of the theory satisfy Poisson bracket versions of the current, or Kac-Moody, algebras,
\EQ{
\big\{\JJ^a_\pm(x),\JJ^b_\pm(y)\big\}&=\epsilon^{abc}\JJ_\pm^c(y)\delta(x-y)\pm\frac {k}{2\pi}\delta^{ab}\delta'(x-y)\ ,\\
\big\{\JJ^a_+(x),\JJ^b_-(y)\big\}&=0\ .
\label{km1}
}
Now if we try to quantize the theory by first regularizing it on a spatial lattice and finding some commutation relations that in some sense follow from \eqref{km1} on the lattice then we run into the severe problem of {\it non ultra-locality\/} arising from the derivative of the delta function in the central term of the classical Poisson bracket algebra. There have been many attempts to solve the problem  directly, whilst maintaining the integrability of the theory at the quantum level, most notably the work of Maillet \cite{Maillet:1985ek,Freidel:1991jv}  (also  \cite{Falceto:1992bf,Melikyan:2016gkd,Delduc:2012vq,Delduc:2012mk,Delduc:2012qb,Benichou:2011ch,Benichou:2010ts,SemenovTianShansky:1995ha} ),  but as yet a universal solution remains elusive.  

Trying to formulate a lattice version of the current algebra is  involved e.g.~\cite{Falceto:1992bf,Alekseev:1992wn}   but also unnecessary. What needs to be done is to find a {\it bare theory\/} on the lattice that lies in the right universality class. Lattice theories that lie in the WZW universality class have been known since the work of Affleck and Haldane \cite{Affleck:1987ch} and Affleck \cite{Affleck:1985wb}, Takhtajan \cite{Takt} and Babujian \cite{Babujian:1982ib}, they are generalized Heisenberg spin chains. In the regularized theory, on a finite lattice the Hilbert space is a product over spin $s$ modules of $\SU(2)$, $V^{\otimes N}$. On each lattice site, the spins satisfy the $\SU(2)$ Lie algebra in a way that is ultra-local across the lattice
\EQ{
[S_m^a,S_n^b]=i\epsilon^{abc}S_m^c\delta_{mn}\ .
}
This algebra, unlike the Kac-Moody algebra \eqref{km1}, lacks the central term proportional to the {\it level\/} $k$. In the bare theory, what becomes the level $k$ in the effective theory in the IR, is  
actually the spin $s=k/2$ of the representation on each lattice site. The KM currents $\JJ^a_\pm$ then arise as renormalized operators which satisfy an operator product version of the KM algebra within correlation functions.  

The beauty of this regularization is that it can maintain integrability for very special chosen Hamiltonians that are constructed by the Quantum Inverse Scattering Method (QISM).\footnote{This method was developed by many authors but there exists an excellent review \cite{Faddeev:1996iy} and a book \cite{KorepinBook} which have many original references.} Once these Hamiltonians are chosen the complete solution is in reach. The spectrum follows from a generalized Bethe Ansatz. The ground state is a non-trivial anti-ferromagnetic state and the physical excitations above it can be found explicitly. The S-matrix of the physical states can then be extracted from the Bethe Ansatz.
The techniques can be extended to correlation functions \cite{KorepinBook}. In general, the spectrum of excitations over the anti-ferromagnetic vacuum state do not have a relativistic dispersion relation because of lattice effects; however, with a carefully chosen continuum limit, what results is a universality class with a relativistic spectrum, S-matrix and correlation functions. There are then various tests and consistency conditions that identify the universality class as the continuum QFT that one was searching for: the WZW and its deformations. 

If the above programme can be completed then we are entitled to say that we have succeeded in quantizing the classical field theory at the most detailed Hilbert space level. Notice that what we have described is a actually a quantum to classical approach, rather than the usual idea of quantizing a classical system. The starting point seems rather removed from the original classical theory that one wished to quantize. The question is whether there is some way of motivating the spin chain directly from the classical integrable theory? Such a construction would be interesting, albeit one that is heuristic because ultimately one would have to quantize the bare theory and take the continuum limit. Then one would have to subject the resulting QFT to certain tests to know that is was describing the original classical theory in a suitable limit. 

A proposal about how one can construct the bare theory directly from the classical theory can be made based on a method proposed by Faddeev and Reshetikhin for investigating the Principal Chiral Model (PCM) via the Quantum Inverse Scattering Method (QISM) \cite{Faddeev:1985qu} but one that can be extended to the WZW models and their deformations \cite{Appadu:2017fff}. In the context of the WZW model, the idea is to consider a deformation of the classical theory at the level of the Poisson brackets \eqref{km1} that involves taking $k\to0$.  This limit is of course rather delicate since (a) it is a very quantum limit applied to classical brackets (recall that $k \rightarrow \infty$ is the semi-classical limit of the WZW) and (b) $k$ is not a continuous parameter.  Nonetheless we continue to take the limit at the level of the classical theory.  At the same time the Hamiltonian of the theory is modified so that the equations of motion, written as a zero curvature conditions for a Lax connection, are preserved. When $k=0$, the Poison brackets \eqref{km1} become ultra local and the classical theory becomes a non-relativistic theory that was called a {\it linear chiral model\/} in \cite{Appadu:2017fff}. This theory can then be regularized on the lattice and then quantized as a spin chain using the light cone lattice construction of Faddeev and Reshetikhin \cite{Faddeev:1985qu} and developed further by Destri and de Vega \cite{Destri:1987ze,Destri:1987hc,Destri:1987ug} (see also the review \cite{Faddeev:1985qu}). This spin chain provides the bare theory which can be solved by Bethe Ansatz techniques. The vacuum state of the relativistic theory is then nothing but the ground state of the spin chain in the anti-ferromagnetic regime and the excitations above it and their S matrix can be found by standard Bethe Ansatz techniques.

\section{Generalized integrable   models}
\label{sec:genint}

In recent years, classical integrable sigma models has been explored in great detail. Starting from the Principal Chiral Models (PCM) and symmetric space models whose integrability has been known for a long time, various kinds of deformations that preserve integrability have been discovered and developed
\cite{Cherednik:1981df,Wiegmann:1985jt,Balog:1993es,Fateev:1996ea,Frolov:2005dj,Klimcik:2002zj,Klimcik:2008eq,Kawaguchi:2011mz,Kawaguchi:2011pf,Lukyanov:2012zt,Delduc:2013fga,Delduc:2013qra,Klimcik:2014bta,Kawaguchi:2014qwa,Hoare:2014pna,Matsumoto:2014nra,Delduc:2014uaa,Hoare:2014oua,Vicedo:2015pna,Matsumoto:2015jja,vanTongeren:2015soa,Hoare:2016hwh,Vicedo:2017cge,Delduc:2017fib,Demulder:2017zhz}. In this work, we will limit attention to theories based on the $\SU(2)$ PCM and in particular to those deformations that preserve one of the global $\SU(2)$ symmetries of the $\SU(2)\times\SU(2)$ symmetry of the PCM,
\EQ{\label{dpcm}
S=-\frac1{2\pi}\int d^2x\,\Tr\big[f^{-1}\partial_+f \BTheta f^{-1}\partial_-f\big]\ ,
}
where $\BTheta$ is endomorphism of the Lie algebra, $\BTheta\cdot T^a=\BTheta_{ab}T^b$. The $\SU(2)$ that is preserved corresponds to left action $f\to Uf$.

The case of $\SU(2)$ is rather special because there are 2 distinct types of integrable deformations. The first are the anisotropic deformations  \cite{Cherednik:1981df,Wiegmann:1985jt,Faddeev:1985qu} which are specific to $\SU(2)$ defined by
\EQ{ \label{eq:aniso_sigma}
\BTheta=\text{diag}(\alpha_1^{-1},\alpha_2^{-1},\alpha_3^{-1})\ .
}
We will mostly be concerned with the XXZ type deformation of this type with $\alpha_1=\alpha_2\equiv\beta$ and $\alpha_3\equiv\alpha$.

The other class, the Yang-Baxter (YB) deformations\footnote{Though outside our main scope the application of these deformations to the $AdS_5 \times S^5$ superstring  \cite{Delduc:2013qra}  has been of considerable interest in holography prompting rich investigation   \cite{vanTongeren:2013gva,Arutyunov:2013ega,Arutynov:2014ota,Ahn:2014aqa,Ahn:2014iia,Ahn:2016egk,Klabbers:2017vtw}.  Notably the spacetime interpretation of these geometry requires a modification to supergravity indicating that the theories are scale but not conformal invariant  \cite{Arutyunov:2015mqj,Wulff:2016tju,Baguet:2016prz,Borsato:2016ose,Araujo:2017enj}.  Further connections have been made in holography to exploit YB deformations in terms of non-commutativity  \cite{Matsumoto:2014gwa,vanTongeren:2015uha,Osten:2016dvf,Hoare:2016wsk,Borsato:2016pas,Hoare:2016wca,Araujo:2017jkb,vanTongeren:2016eeb}.} found by Klimcik \cite{Klimcik:2002zj,Klimcik:2008eq}, exist for any Lie group and take the form
\EQ{
\BTheta=\alpha^{-1}(1-\eta\BR )^{-1}\ ,
}
where $\BR$ is any solution of the (modified) classical YB equation
\EQ{
[\BR a,\BR b]-\BR\big([\BR a,b]+[a,\BR b]\big) =-c^2[a,b]\ , 
\label{YB1}
}
for all $a,b$ in the Lie algebra and where $c^2$, modulo scalings being $\in \{ -1, 1,0\}$. For $\SU(2)$ there is a single deformation of this type with $c^2=-1$ given by
\EQ{
\BR _{ab}=\MAT{0 & -1 & 0\\ 1 & 0 & 0\\ 0 & 0 & 0}\  .
}
Note that the YB and anisotropic XXZ deformations only differ by a total derivative, which vanishes if space is compact with parameters related via
\EQ{\label{abtoetat}
1+\eta^2= \frac\beta\alpha\ .
}
The parameter $\eta$ is real and so the YB model can only be equivalent to the anisotropic model in a compact space if $\beta>\alpha$.

If we compute the Poisson brackets of these theories then they do not take the form of the current algebras \eqref{km1}. So it seems that we cannot quantize this theory by using the spin chain scenario outlined in the introduction; indeed, at the quantum level, we know that the PCM itself is asymptotically free and not a WZW model. 

There was a suggestion in the early literature by Polyakov and Wiegman \cite{Polyakov:1983tt} that the PCM could -- -in some senses -- -be recovered via a $k\to\infty$ limit of the WZW model. This idea was more recently understood as an example of ``regularized" non-abelian T-duality. This latter idea comes from string theory but can be understood as a procedure can be applied to sigma model with a global symmetry $G$. Non-abelian T duality applied to this symmetry leads to theory with non-compact directions in field space  \cite{delaOssa:1992vci,Alvarez:1993qi,Giveon:1993ai,Alvarez:1994np,Sfetsos:2010uq}. A procedure for completing, or regularizing, this geometry to arrive at a dual theory, the ``lambda model", with a compact geometry naturally leads to a deformed WZW model  \cite{Evans:1994hi,Sfetsos:2013wia}. The sigma model is recovered in the large level limit $k\to\infty$ with an IRF-to-vertex transformation.
 
The most efficient way to construct the lambda model, is to take the original sigma model with global $G$ symmetry, acting, say for left multiplication on the group field $f\to Uf$, and add to it a decoupled WZW model also for the group $G$, with field $\CF$ and action $\CF\to U\CF U^{-1}$, and then gauge the common $G$ action \cite{Sfetsos:2013wia}. If the sigma model is a PCM the now coupled theory can be gauge fixed by choosing $f=1$. What remains is the WZW model with the whole $G$ vector action gauged but with a deformation:
\EQ{  
S=k\, S_\text{gWZW}[\CF,A_\mu]-\frac1{2\pi}\int d^2x\,\Tr\big[A_+\BTheta A_-\big]\ .
\label{sma}
}
Although at first sight if appears that the $G$ action is gauged, in the presence of the deformation this is false and one should actually view the field $A_\mu$ as just an auxiliary field that can be integrating out exactly at Gaussian order. If we perform this integral then what remains is a current-current type deformation of a WZW model:
\EQ{\label{eq:act}
S=k\,S_\text{WZW}[{\cal F}] + \frac{k}{2\pi} \int d^2 \sigma \Tr\left( {\cal F}^{-1} \partial_+ {\cal F} (\BOmega  - {\textrm{Ad}}_{\cal F})^{-1} \partial_- {\cal F} {\cal F}^{-1} \right)  \ ,
}
with
\EQ{
\BOmega=\BId+k^{-1}\BTheta\ .
}
Deformations of this type has been extensively studied in the literature 
  \cite{Itsios:2014lca,Sfetsos:2014jfa,Hollowood:2014rla,Itsios:2014vfa,Hollowood:2014qma,Sfetsos:2014cea,Sfetsos:2014lla,Demulder:2015lva,Hoare:2015gda,Sfetsos:2015nya,Hollowood:2015dpa,Appadu:2015nfa,Klimcik:2015gba,Georgiou:2015nka,Borsato:2016zcf,Georgiou:2016iom,Chervonyi:2016ajp,Klimcik:2016rov,Borsato:2016ose,Chervonyi:2016bfl,Schmidtt:2016tkx,Georgiou:2016zyo,Georgiou:2016urf,Schmidtt:2017ngw,Georgiou:2017aei,Appadu:2017fff,Appadu:2017xku,Georgiou:2017oly,Borsato:2017qsx,Appadu:2017bnv,Georgiou:2017jfi,Roychowdhury:2017vdo,Sfetsos:2017sep,Lunin:2017bha}. For the $\SU(2)$ anisotropic and YB models considered here, see \cite{Sfetsos:2014lla,Appadu:2017bnv}.

The Poisson brackets of the lambda models is precisely the current algebra \eqref{km1} (before integrating out the auxiliary field $A_\mu$) where the currents take the form
\EQ{
\JJ_+&=-\frac k{2\pi}\big(\CF ^{-1}\partial_+\CF +\CF ^{-1}A_+\CF -A_-\big)\ ,\\
\JJ_-&=\frac k{2\pi}\big(\partial_-\CF \CF ^{-1}-\CF A_-\CF ^{-1}+A_+\big)\ .
\label{ksq2}
}
Therefore it should be possible to quantize the lambda models in the Hamiltonian formalism on a lattice as outlined above by identifying the correct Hamiltonian. The associated sigma models would then be recovered by a $k\to\infty$ limit  and an IRF-to-vertex transformation\footnote{This is a transformation on the Hilbert space of lattice models originally  introduced by Baxter \cite{Baxter:1972wf} in the context of the 8-vertex model.  Lattice variables  assigned to lattice sites in the face model (or SOS model) become reassigned to edges in a vertex model and vice versa.  It seems plausible that this is the way that non-Abelian T-duality is manifested at the level of the Hilbert space, though one should like to make this connection more precise.}  on the physical states \cite{Appadu:2017fff}. This latter transformation amounts to a change of basis in the Hilbert space from a kink picture of the Hilbert space to that of local excitations  \cite{LeClair:1989wy,Bernard:1990ys,Bernard:1990cw}.

\vspace{0.2cm}
\noindent{\bf The XXZ lambda model}

In this work, we will focus on the anisotropic XXZ lambda model for which
\EQ{\label{eq:OmegaXXZ}
\BOmega=\MAT{\xi^{-1}& 0 & 0\\ 0 & \xi^{-1} & 0\\ 0 & 0 & \lambda^{-1}}\ ,
}
where the parameters are related to those of the anisotropic sigma model (c.f. eq.~\eqref{eq:aniso_sigma}) via    
\EQ{
\xi=\frac{k\beta}{k\beta+1}\ ,\qquad\lambda=\frac{k\alpha}{k\alpha+1}\ .
\label{zas}
}
Note that the YB lambda model is not equivalent to the XXZ lambda model even in a compact space; for instance, the former breaks parity  \cite{Appadu:2017bnv}.

At this point, it is worth pointing out that in the limit $\lambda\to1$, the abelian component of the gauge field $\Tr(T^3A_\mu)$ becomes a genuine gauge symmetry and the theory reduces to the symmetric space $\SU(2)/\U(1)$ lambda model (with its lambda parameter equal to $\xi$).  This is  apparent when viewing the lambda theory prior to integrating out gauge fields as in eq.~\eqref{sma}; here the same limit consists of sending $\alpha^{-1} \rightarrow 0$ such that $\BTheta  \rightarrow {\textrm{diag}}(\beta^{-1} , \beta^{-1} ,0 )$  hence removing all symmetry breaking terms associated $\Tr(T^3A_\mu)$.   Indeed the same limit applied to the PCM of eq.~\eqref{dpcm} serves to construct the $\text{O}(3)$ model i.e.~the sigma model with an $S^2$ target space \cite{Karowski:1978iv}.

The XXZ lambda model is classical integrable which is proved by writing the equations of motion in Lax form. To start with the equations of motion of the auxiliary field $A_\mu$ are second class constraints
\EQ{
\CF^{-1}\partial_+\CF+\CF^{-1}A_+\CF&=\BOmega^TA_+\ ,\\
-\partial_-\CF\CF^{-1}+\CF A_-\CF^{-1}&=\BOmega A_-\ .
}
Imposing these constraints, the equations of motion of the group valued field $\CF$ can be written in terms of a Lax connection valued in the loop algebra of $\msu(2)$:\footnote{The loop variable is $e^z$, and the algebra has the principal gradation: $T^a e^{(2n-1)z}$, $a=1,2$, and  $T^3e^{2nz}$. Here, and in the following, we use a basis $\{T^a\}$ that are anti-hermitian and normalized so that $\Tr(T^aT^b)=-\delta^{ab}$. }
\EQ{ \label{eq:lax}
\Lax_\pm(z)=\sum_{a=1}^3w_a(\nu\mp z)A^a_\pm T^a\ ,
}
with
\EQ{
w_1(z)=w_2(z)=\sqrt{\frac{\lambda^2-\xi^2}{1-\lambda^2}}\cdot\frac1{\xi\sinh z}\ ,\qquad
w_3(z)=\sqrt{\frac{\lambda^2-\xi^2}{1-\xi^2}}\cdot\frac1{\lambda\tanh z} \ ,
}
and
\EQ{
\cosh^2\nu=\frac{(1-\xi)(\lambda+\xi)}{2\xi(1-\lambda)}\ .
\label{inh}
}
In the sigma model limit $k\to\infty$, the Lax equations reduce to the equations of motion of the XXZ sigma with $A_\pm$ identified with $f^{-1}\partial_\pm f$.

It was shown in \cite{Appadu:2017bnv} that the XXZ lambda model has an affine quantum group symmetry  $\mathscr U_q(\widehat{\msu(2)})$ (in the principal gradation) at the classical level where the deformation parameter is 
\EQ{
q=\exp\big[-i\pi/\gamma'\big]\ ,\qquad \gamma^{\prime2}=\frac{k^2}4\cdot\frac{(1-\xi^2)(1-\lambda)^2}{\lambda^2-\xi^2}\ .
\label{u7a}
}
Note that the classical limit corresponds to $k\to\infty$ but this can be taken keeping $q$ fixed so that the quantum group algebra becomes realized at the Poisson bracket level. The model also has another  $\mathscr U_q(\widehat{\msu(2)})$  affine quantum group symmetry, but with
\EQ{
q=\exp\big[-i\pi/(k+2)\big]\ .
}
This leaves a Yangian symmetry in the classical limit. 

The quantum behaviour of the XXZ lambda model depends crucially on whether $\lambda\lessgtr\xi$, i.e.~on whether $q$ is real ($\lambda<\xi$) or $q$ is a complex phase ($\lambda>\xi$). The behaviour becomes apparent in the RG flow  \cite{Sfetsos:2014jfa}
\EQ{
\mu\frac{d\xi}{d\mu}&=\frac4k\cdot\frac{\xi(\xi^2-\lambda)}{(1-\xi^2)(\lambda+1)}+{\cal O}(1/k^2)\ ,\\ \mu\frac{d\lambda}{d\mu}&=-\frac 4k\cdot\frac{\xi^2(1-\lambda)^2}{(1-\xi^2)^2}+{\cal O}(1/k^2)\ .
\label{rg2}
}  
It is important that the deformation parameter \eqref{u7a} is an RG invariant. When $\gamma'$ is real ($q$ is a complex phase), the RG flow is UV safe with $\xi\to0$ and $\lambda\to\lambda_*$, 
\EQ{
\lambda_*=\frac k{2\gamma'+k}
}
a constant, in the UV. We call this the ``UV safe XXZ lambda model". We can use the RG invariant to write a closed beta function equation for $\lambda$ at the one-loop level:
\EQ{
\mu\frac{d\lambda}{d\mu}=\DT{-} \frac 4k\cdot\frac{\big(2\lambda_*-1+\lambda_*^2(\lambda-2)\lambda\big)\big(\lambda_*^2-2\lambda_*^2\lambda+(2\lambda_*-1)\lambda^2\big)}{(\lambda_*-1)^4(1+\lambda)^2}+{\cal O}(1/k^2)\ .
\label{rg3}
}

On the contrary, when $\gamma'=i\sigma'$ is imaginary ($q$ is real) the RG flow is cyclic \cite{Appadu:2017bnv}. We call this the ``cyclic XXZ lambda model". It is far from clear what the full implications of a cyclic RG behaviour are  \cite{Mussardo:1999ee,Leclair:2003xj,LeClair:2003hj,LeClair:2004ps}.

In  \cite{Appadu:2017bnv}, on the basis of the symmetries of the theories, we proposed factorizable S-matrices to describe quantum XXZ lambda models. In the UV safe regime, the S-matrix takes the block form 
\be
S = S_\text{SG}(\theta; \gamma' ) \otimes S_\text{RSOS}(\theta;k)\ ,
\label{sm8}
\ee
where $S_\text{SG}$ is the sine-Gordon soliton S-matrix \cite{Zamolodchikov:1978xm} and $S_\text{RSOS}$ is the restricted sine-Gordon kink S-matrix \cite{LeClair:1989wy}. The notation here, implies that states transform in a product of quantum numbers of each of the two factors and the S-matrix does not mix up the quantum numbers of each factor. The S-matrix \eqref{sm8} is known as the ``fractional supersymmetric sine-Gordon S-matrix" \cite{Bernard:1990ti}.

In the cyclic regime, the conjectured S-matrix was  \cite{Appadu:2017bnv}
\EQ{
S = S_\text{cyclic-SG}(\theta; \sigma' ) \otimes S_\text{RSOS}(\theta;k)\ ,
\label{sm9}
}
where the first block is the cyclic sine-Gordon S-matrix which is a close cousin of the sine-Gordon soliton S-matrix, but with different analytic structure so that the S-matrix is periodic in rapidity shifts $\theta\to\theta+\pi\sigma'$  \cite{Leclair:2003xj}.\footnote{In fact the S-matrix changes sign under the shift, but this sign change is not observable.} It is important to note that the complete S-matrix is not periodic in rapidity due to the RSOS piece. However, periodicity is restored at high energy where the $S_\text{RSOS}$ piece becomes trivial. On the other hand, in the $k\to\infty$ limit plus an IRF-to-vertex transformation, the RSOS piece becomes the $\SU(2)$ invariant S-matrix block that reflects the symmetry of the $\SU(2)_L$ symmetry of the XXZ sigma model.

It is also worth saying that the S-matrix of the YB lambda model is also of the same form \eqref{sm9} but with the affine quantum group representation in a different gradation: homogeneous rather than principal. This change in gradation reflects the fact that parity is broken in the YB lambda model \cite{Appadu:2017bnv}.

\section{Spin Chain via the FR Limit}
\label{sec:FR}
In this section, we describe the, partly heuristic, Faddeev-Reshetikhin (FR) procedure \cite{Faddeev:1985qu} that leads to identifying a spin chain associated to the XXZ lambda model. The idea is to take the classical theory and consider a limit where $k\to0$ and 
$\xi,\lambda\to0$ keeping the ratios $k/\xi$ and $k/\lambda$ fixed. In this limit, it is clear from the Kac-Moody Poisson brackets \eqref{km1}, that the resulting theory becomes ultra local since the central terms involving $\delta'(x-y)$ disappear. In the FR limit, the Lax connection of eq.~\eqref{eq:lax}
 becomes
\EQ{
\Lax_\pm(x;z)=- \frac{\pi}{\gamma'\sinh(\nu\mp z)}\Big(\JJ_\pm^1T^1+\JJ_\pm^2 T^2+\cosh(\nu\mp z)\JJ_\pm^3T^3\Big)\ , 
\label{lmm}
}
with
\EQ{
\cosh^2\nu=\frac{\lambda+\xi}{2\xi}\ .
}

The theory in the FR limit can be written as a non-relativistic field theory, the {\it linear chiral model\/} \cite{Appadu:2017fff}. But here we proceed more directly to our goal of constructing the spin chain. There are two steps that we combine. The first is to regularize the theory on a spatial lattice, whilst preserving integrability, and then quantize the discrete modes of the currents $\JJ^a_\pm$, which is straightforward because the Poisson bracket is ultra local. 

The most elegant way to describe the regularized theory is via a light-cone lattice in 2d Minkowski spacetime \cite{Destri:1987ze,Destri:1987hc,Destri:1987ug}
\EQ{
x^+=t+x=nL\ ,\qquad x^-=t-x=mL\ ,\qquad m,n\in\mathbb Z\ .
}
Taking $t=0$, we define the discretized currents $\JJ^a_{\pm,n}=\JJ^a_\pm(x_n)$ with $x_n=nL$, which naturally lie on the null links of the lattice at $t=0$. It is now simple to quantize the model, by replacing the discretrized Poisson bracket algebra by commutators:
\EQ{
&\{\JJ_\pm^a(x),\JJ^b_\pm(y)\}=\epsilon^{abc}\JJ^c_\pm(x)\delta(x-y)\\ &\longrightarrow\qquad \{\JJ_{\pm,m}^a,\JJ^b_{\pm,n}\}=\frac1L\epsilon^{abc}\JJ^c_{\pm,m}\delta_{mn}\\
&\longrightarrow\qquad [\JJ_{\pm,m}^a,\JJ^b_{\pm,n}]=\frac iL\epsilon^{abc}\JJ^c_{\pm,m}\delta_{mn}\ .
}
So the discrete currents satisfy the $\msu(2)$ Lie algebra and can be quantized as operators on a spin $k/2$ representation:\footnote{The operators $S^a$ are formally identical to the generators $iT^a$ in a spin $s$ representation.}
\EQ{
\JJ_{+,n}^a=\frac 1L S_{2n}^a\ ,\qquad \JJ_{-,n}^a=\frac 1L S^a_{2n-1}\ .
\label{fij}
} 
Note that we define the spin via $k$ because it will transpire that this is where the WZW level, lost in the FR limit, will reappear. So the Hilbert space is a product $V_1\otimes V_2\otimes\cdots \otimes V_{2p}$, where $V$ is the $k+1$-dimensional module for the spin $k/2$ representation, and so the quantum model is a type of higher spin Heisenberg spin chain. 

Of particular importance for establishing the integrability of the quantum model is to identify the analogue of the Lax connection. This will allow us to deploy the powerful machinery of the QISM including the Algebraic Bethe Ansatz in order to solve the model. The natural object to consider is the monodromy of the Lax connection, which in the discrete model involve the parallel transport, or Wilson line, along the null links viewed as quantum operators  \cite{Faddeev:1985qu,Destri:1987ze,Destri:1987hc,Destri:1987ug}:
\EQ{
&\text{P}\overset{\longleftarrow}{\text{exp}}\Big[-\int_{nL}^{(n+1)L}dx^+\,\Lax_+(x;z)\Big] \longrightarrow R_{2n-1,0}(z-\nu)\ ,\\
&\text{P}\overset{\longleftarrow}{\text{exp}}\Big[-\int_{nL}^{(n-1)L}dx^-\,\Lax_-(x;z)\Big] \longrightarrow R_{2n,0}(z+\nu)\ .
\label{rnn2}
}
The object $R_{n,0}(z)$ acts as a map on a product of $\msu(2)$ representations $V_n\otimes V_0\to V_n\otimes V_0$, where $V_n$ is a quantum space, part of the Hilbert space, and $V_0$ is an auxiliary space that handles the fact that the Lax connection lies in the $\msu(2)$ Lie algebra. 
The shifted dependance on the spectral parameter $z$ in $R_{2n-1,0}(z-\nu)$ and $R_{2n,0}(z+\nu)$  simply reflects that of the Lax connection eq.~\eqref{lmm} and will have an important interpretation in our spin-chain picture as the inclusion of alternating inhomogeneities.

In order to be able to define a suitable Hamiltonian for the spin chain it is necessary that the auxiliary space is the same spin $k/2$ representation as those that build the Hilbert space so that we have the {\it regularity\/} condition
\EQ{\label{eq:R0}
R(0)=P\ ,
}
the permutation of the two spin $k/2$ spaces.\footnote{This is explained in Faddeev's review \cite{Faddeev:1996iy}.}
The key to integrability is to demand that $R_{n,0}(z)$ satisfies the Yang-Baxter equation
\EQ{
R_{0,0'}(z-w)R_{n,0}(z)R_{n,0'}(w)=R_{n,0'}(w)R_{n,0}(z)R_{0,0'}(z-w)\ ,
\label{ybe2}
}
where we have indicated the auxiliary spaces as $0$ and $0'$. 
Moreover, we require that in the classical and continuum limit that
\EQ{
R_{n,0}(z\mp \nu)\longrightarrow 1\mp L\Lax_\pm(x_n,z)+\cdots\ ,
\label{fii}
}
where the classical Lax connection is \eqref{lmm}. These conditions then completely determine the $R$-matrix completely and identify it as the well-known trigonometric $R$-matrix associated to $\msu(2)$ for the spin $k/2$ representation.
The construction of this $R$-matrix is somewhat involved but can be obtained from that of the spin $\frac{1}{2}$ representation via an analogue of the way  irreducible representations are arrived at by   decomposing tensor products of the fundamental \cite{Kulish:1981gi}.\footnote{At an algebraic level one can follow a ``descent''  procedure starting with a solution $T^{ \Lambda}(u)$ of the RTT relation eq.~\eqref{fcr} in a representation $\Lambda$ with the auxiliary space still the fundamental.  Then one can search for $R$-matrices $R^{(\Lambda_1,\Lambda_2)}$ that obey $T^{ \Lambda_1}(u)_{12} T^{ \Lambda_2}(u)_{13} R^{(\Lambda_1,\Lambda_2)}_{23} (u-v) =  R^{(\Lambda_1,\Lambda_2)}_{23} (u-v) T^{ \Lambda_2}(u)_{13} T^{ \Lambda_1}(v)_{12} $.  In this way a collection of R-matrices in various representations are built that obey YB equations $ R^{(\Lambda_1,\Lambda_2)}_{12} (u)R^{(\Lambda_1,\Lambda_3)}_{13} (u+v)R^{(\Lambda_2,\Lambda_3)}_{23} (v) = R^{(\Lambda_2,\Lambda_3)}_{23} (v)R^{(\Lambda_1,\Lambda_3)}_{13} (u+v) R^{(\Lambda_1,\Lambda_2)}_{12} (u)$.   See section 8 of \cite{Faddeev:1996iy} for a pedagogical treatment in the case of the $XXX_s$ spin chain.}  Here the $R$-matrix in the fundamental,  $R^{(\frac{1}{2},\frac{1}{2})}$,  is given by
\EQ{\label{eq:rmat}
R^{(\frac{1}{2},\frac{1}{2})}(z)&=(e^zq-e^{-z}q^{-1})(e_{11}\otimes e_{11}+e_{22}\otimes e_{22})\\ &
+(q-q^{-1})(e_{12}\otimes e_{21}+e_{21}\otimes e_{12})+(e^z-e^{-z})(e_{11}\otimes e_{22}+e_{22}\otimes e_{11})\ ,
}
where $e_{ij}$ is a $2\times2$ matrix with a 1 in position $ij$ and 0's elsewhere. 
The $R$-matrix depends on the deformation parameter $q=\exp[-\gamma]$ such that the classical limit involves taking $\gamma\to0$, in which case
\EQ{
cR(z)&\overset{\gamma\to0}{\longrightarrow}1+\frac\gamma{i\sinh z}\big(T^1\otimes T^1+T^2\otimes T^2+\cosh z\, T^3\otimes T^3\big)\\
&\equiv1+\frac\gamma{\sinh z}\big(S^1T^1+S^2T^2+\cosh z\, S^3T^3\big)\ .
}
where $c$ is a suitable normalization factor.
Note that for the spin $\frac12$ representation ($k=1)$ this expression is the exact expression for the trigonometric $R$-matrix. 

Then given \eqref{fij}, \eqref{fii} we can precisely extract the classical Lax connection\eqref{lmm} with the identification
\EQ{
\gamma' \longleftrightarrow \frac\pi\gamma\ .
\label{rgg}
}
Note this is only valid in the FR limit ($k\to0$) and we will find a finite $k$ correction later.

In order to define the spin chain, we have to define the energy and momentum operators acting on the Hilbert space. This is where the QISM comes to the fore. In the QISM the key quantity is the monodromy matrix for the whole lattice obtained by taking a product over the null links of the lattice at a given time step \eqref{rnn} 
\EQ{
T(z)=R_{N,0}(z+\nu)R_{N-1,0}(z-\nu)\cdots R_{1,0}(z-\nu)\ .
\label{juw}
}
This is the monodromy matrix of an XXZ spin chain with alternating inhomogeneities $(-1)^n\nu$ and anisotropy parameter $\gamma$. 
The key to the QISM is that by virtue of the Yang-Baxter equation the monodromy matrix satisfies generalized commutation relations of the form
\EQ{
R_{00'}(z-y)T_0(z)T_{0'}(y)=T_{0'}(y)T_0(z)R_{00'}(z-y)\ .
\label{fcr}
}
where we have indicated the auxiliary spaces as $0$ and $0'$. It follows that the trace of the monodromy matrix on the auxiliary space provide a class of commutating operators on the Hilbert space:
\EQ{
\big[\Tr_0 T(z),\Tr_0 T(y)]=0\ .
}

In the light cone lattice approach, the null components of the energy and momentum are associated to the shift operators \cite{Faddeev:1985qu,Destri:1987hc} 
\EQ{
U_+=e^{-iL{\cal P}_+}=\Tr_0T(\nu)\ ,\qquad U_-^\dagger=e^{iL{\cal P}_-}=\Tr_0T(-\nu)\ .
}
These unitary operators generate shifts on the light cone lattice $x^+\to x^++L$ and $x^-\to x^ -- L$.\footnote{A detailed proof of this is found in section 11 of the review \cite{Faddeev:1996iy}, but the essential point is that  when $z = \pm \nu$ half of the $R$-matrices entering in the monodromy  reduce to permutations via eq.~\eqref{eq:R0}.}  Note that $U_+$ commutes with $U_-$ on account of \eqref{fcr}.
We can therefore express the energy (Hamiltonian) and momentum in terms of the trace of the monodromy matrix
\EQ{
{\cal E}&\equiv H = {\cal P}_++{\cal P}_-  =\frac iL\log\big[\Tr_0T(\nu)/\Tr_0 T(-\nu)\big]\ ,\\{\cal P}&= {\cal P}_+-{\cal P}_-  = \frac iL\log\big[\Tr_0 T(\nu)\Tr_0 T(-\nu)\big]\ .
\label{hpp}
}
Note that the energy operator is not local on the spin chain as it is for the standard Heisenberg chain.

The expressions \eqref{hpp} are the starting point for applying the QISM to our theory. Before we turn to this problem, it is useful to note that the light-cone lattice formulation yields directly the Lax equation of motion in the classical and continuum limit \cite{Faddeev:1996iy}. On the lattice one has
\EQ{
R_{2n,0}(z+\nu)R_{2n-1,0}(z-\nu)=U_+^{-1}R_{2n-1,0}(z-\nu)U_+U_-^{-1}R_{2n,0}(z+\nu)U_-\ ,
}
which, on taking the classical continuum limit becomes the flatness condition
\EQ{
\partial_+\Lax_-(z)-\partial_-\Lax_+(z)+[\Lax_+(z),\Lax_-(z)]=0\ .
}
 
\section{The Spin Chain}
\label{sec:spinchain}

In this section, we discuss the spin chain that was constructed in the last section that we want to identify with the XXZ lambda model. This spin chain is a higher spin version of the well-known XXZ Heisenberg spin chain  but with a novel Hamiltonian.\footnote{The XXZ $s=\frac12$ Heisenberg spin chain is discussed in the books by Takahashi \cite{Tak} and Sutherland \cite{Suth}. The Bethe Ansatz equations are analysed in detail in \cite{Babelon:1982mc}. The higher spin XXZ chain is analysed in \cite{Sogo,Kirillov:1987zz,Frahm:1990dwx}.} To put some flesh on the bones, we begin our discussion with the conventional higher spin XXZ Heisenberg spin chain which is described by a Hamiltonian
\EQ{
H_\text{XXZ} &=-i\frac d{dz}\log \Tr_0T(z)\Big|_{z=0}+\text{const.}\\ &= \sum_{n=1}^{N-1} \Big(S^1_nS^1_{n+1} +S^2_nS^2_{n+1} -\Delta\, S^3_n S^3_{n+1} \Big)+\big(\text{higher terms}\big)\ ,
\label{ham1}
}
where the monodromy matrix here is identified with \eqref{juw} but with vanishing inhomogeneity parameter $\nu=0$. The higher terms are polynomials in $S_n^aS^a_{n+1}$ of degree $2s$.
The anisoptropy parameter here is
\EQ{
\Delta=-\cos\gamma\ ,
}
where $\gamma$ appeared earlier in the $R$ matrix. The sign of the XXZ Hamiltonian can be flipped by conjugation by a simple operator which also sends $\Delta\to-\Delta$: $H_\text{XXZ}(\Delta)=-JH_\text{XXZ}(-\Delta)J$.

It is important to emphasize that $H_\text{XXZ}$, the Hamiltonian of the Heisenberg chain, is not the Hamiltonian of the light-cone lattice model (which, as we have remarked, is not local), but, as we shall see, they share the same eigenstates. For the Heisenberg chain the model has several phases depending on the value of $\Delta$.
When $\Delta\geq1$, the system is ferromagnetic, with $\Delta=1$ being the conventional ferromagnetic XXX Heisenberg spin chain. The ``easy plane" regime $|\Delta|<1$ is paramagnetic and gapless. When $\Delta=-1$, the model is the (gapless) anti-ferromagnetic XXX Heisenberg spin chain.  Finally there is a regime with $\Delta<-1$ which is anti-ferromagnetic with a gap. In this regime, we will write
\EQ{
\gamma=i\sigma\ ,\qquad \sigma\in\mathbb R\ ,
}
hence, $\Delta=-\cosh\sigma$.
Note that with the relation between $\gamma$ and $\gamma'$ in \eqref{rgg}, the ``UV safe regime" of the lambda model maps to the easy plane region $1>|\Delta|$, while the ``cyclic regime" maps to the gapped anti-ferromagnetic regime $\Delta<-1$. This is important for our goal of using the spin chains to quantize relativistic field theories because the existence of a continuum limit relies on working in a gapless regime where the correlation length is infinite in lattice units allowing a limit where the lattice spacing goes to zero keeping a finite correlation length fixed. We will be returning to this important point as we proceed.

It is not important to write out the explicit form of the Hamiltonian, which would be complicated, but rather we want to apply the algebraic Bethe Ansatz to find the spectrum. This will allow us to find the ground state in the thermodynamic limit and the spectrum of excitations and their $S$-matrix over the ground state. We will not pursue a complete analysis here, rather our intention is to provide enough detail so that we can identify the physical $S$-matrix of the theory in terms of a known factorizable $S$-matrix and the show when a continuum limit exists that leads to a relativistic integrable QFT, to be identified, with the XXZ lambda model.

The eigenstates of the XXZ spin chain and the light-cone model are both equal to the eigenstates of the trace of the monodromy matrix $\Tr_0T(z)$ are these can be constructed using the algebraic Bethe Ansatz. With alternating inhomogeneities $\pm\nu$, they can be written 
\be
\exp\big[iN\theta_{2s}(z_i+\nu)/2+iN\theta_{2s}(z_i-\nu)/2\big] =- \prod_j \exp\big[i\theta_2(z_i-z_j)\big]\ ,
\label{BAE}
\ee
where we define the function 
\EQ{
e^{i\theta_n(z)} =\frac{\sinh(z-in\gamma/2) }{\sinh(z+in\gamma/2) } \qquad (1>|\Delta|)\ ,\qquad
\frac{\sin(z-in\sigma/2)}{\sin(z+in\sigma/2)} \qquad (\Delta<-1)\ .
}
For the conventional spin chain one sets $\nu=0$.

The set $\{z_i\}$, $i=1,2,\ldots,m$, are the Bethe roots. Every set of Bethe roots labels an eigenstate of the monodromy  matrix $\Tr_0T(z)$. In particular, the eigenvalues of the null evolution operators are then
\EQ{
U_\pm=\exp\Big[\pm i\sum_{j=1}^m\theta_{2s}(z_j\pm\nu)\Big]\  .
}
Hence, the energy and momentum can be written as 
\EQ{
{\cal E}=\sum_{j=1}^m\varepsilon(z_j)\ ,\qquad {\cal P}=\sum_{j=1}^m\wp(z_j)\ .
}
So each Bethe root $z_j$ can be interpreted as a pseudo particle carrying energy and momentum 
\EQ{
\varepsilon(z)&=\frac 1{2L}\big(\theta_{2s}(z+\nu)-\theta_{2s}(z-\nu)\big)\ ,\\
\wp(z)&=\frac1{2L}\big(\theta_{2s}(z+\nu)+\theta_{2s}(z-\nu)\big)\ ,\label{ppe}
}
with the branches of the logs in the definition of $\theta_n(z)$ chosen so that $\wp(0)=0$ and
\EQ{
\varepsilon(0)=\frac2L\begin{cases}\arctan(\cot(s\gamma)\tanh(\nu))-\pi\ , & 1>|\Delta|\ ,\\
\arctan(\coth(s\sigma)\tan(\nu))-\pi\ , & \Delta<-1\ .\end{cases}
}
Notice that the variable $z$ is a kind of rapidity variable for the pseudo particle. The energy of the pseudo particles is negative for all rapidities and so we can expect the vacuum state to be a Dirac sea of pseudo particles. However, it  transpires that the pseudo particles form patterns known as strings.

The string hypothesis proposes that the dominant contribution to the chain  in the thermodynamic limit take the form of $n$-{\em strings} -- -collections of $n$ Bethe roots sharing a common real part with equally distributed  imaginary parts    such that the total momentum of the string is real. In the higher spin XXZ model such strings are classified by their length $n$ and their parity $\epsilon = \pm 1$, as well as their centre $z$. In the $1>\Delta>-1$ regime \cite{Kirillov:1987zz}
\be
z_j=  z + \frac{i\gamma}2\Big(  n+1 -2 j  + \frac{\pi}{2\gamma} (1-\epsilon_s  \epsilon)\Big)  \ , \qquad j = 1,2, \ldots, n \ , 
\ee
in which $\epsilon_s= (-1)^{[2s\gamma/\pi] }$. However, the strings can not be chosen arbitrarily; for a given coupling $\gamma$ the allowed values of $\{n_i\}$ and parity $\{ \nu_i\}$  are constrained by inequalities 
\be
\nu_i \sin(\gamma(n_i-j))\sin(\gamma j) > 0 \ ,\qquad j = 1 ,2, \ldots,n_i -1 \ . 
\ee
The solutions $n_i$ to these conditions  inequalities are known as Takahashi numbers, a set of integers that can be extracted from the continued fraction expansion expansion of $\pi/\gamma$ (for details see \cite{Frahm:1990dwx} where $\pi/\gamma=p_0$).

From an analysis of the Bethe ansatz equation Kirillov and Reshetikhin \cite{Kirillov:1987zz} found a further requirement that the spin $s$ and the anisotropy $\gamma$ should be related such that $2s+1$ is exactly one of the admissible string lengths.   It was later shown by Frahm et al \cite{Frahm:1990dwx} that this relation meshes precisely with the requirement that the conventional XXZ spin chain Hamilitonian \eqref{ham1} is hermitian.   For a given value of spin $s$ one finds that the set of admissible values of $\pi/\gamma$ form a collection of disjoint intervals.  In the thermodynamic limit, the contributions to the ground state consists of a  {\em Dirac sea\/}  of negative energy strings, positive energy strings {\em breathers\/}  and the remaining strings of zero energy.  Both the breather modes and holes in the Dirac sea should be viewed as excitations over the ground state. We will focus initially on the holes and then discover the breather via the bootstrap of the physical S-matrix.

The structure of the ground state is simpler in the regime $2s<\pi/\gamma$ relevant for us, for which the ground state Dirac sea is filled by one type of string of length $2s$ and parity $\epsilon = 1$. This picture also extends into the $\Delta\geq 1$ regime. So the $2s$ string in the two regimes of interest has the simple form
\EQ{
z_j=z+\begin{cases}\frac{i\gamma}2(1+2s-2j)\ ,    \qquad & 1>|\Delta|\ ,\\ \frac{i\sigma}2(1+2s-2j)\ ,\qquad & \Delta<-1\ ,\end{cases}\qquad j=1,2,\ldots,2s\ .
}
From now on, we will focus on $2s$ strings and their holes.
The scattering of two $2s$ strings with centres $z_1-z_2\equiv z$ is described by scattering their constituent parts, i.e. 
\EQ{
K_{2s,2s}(z)=i\frac d{dz}\log S_{2s,2s} (z) = \sum_{j,l=1}^{2s} \theta'_2(z_{1,j}-z_{2,l}) =  \theta'_{4s}(z) +\sum_{j=1}^{2s-1}2\theta'_{2j}(z)  \ . 
}
We will also need to define the kernel
\EQ{
\theta'_{2s,2s}(z)=\sum_{j=1}^{2s } \theta'_{2s}(z_j) =\sum_{j=1}^{2s}\theta'_{2j-1}(z)\ .
}
 
The next step is to take the thermodynamic limit, that is $N\to\infty$, in which case the distribution of Bethe roots and holes are described by densities (unit normalized). In the present context, we are not going to undertake a full analysis of all the excitations over the physical ground state described by the Thermodynamic Bethe Ansatz (TBA). Rather, in order to identify the physical excitations and their S-matrix we only need a truncated version of the analysis of the TBA in which we only consider the excitations characterised by a density of holes $\rho_h(z)$ holes in a Dirac sea of such length $2s$-strings with density density $\rho(z)$. In addition, we will not concern ourselves with boundary conditions. The $2s$ string hole excitations -- -the spinons -- -are highest weight states of the (quantum group) symmetries of the spin chain, and knowledge of their structure and scattering allows us to exact the full spectrum and S-matrix. 

The TBA equation is obtained by taking the the thermodynamic limit of \eqref{BAE}:
\be
\rho(z) + \rho_h(z) +K_{2s,2s}\ast\rho(z)=\frac1{4\pi}\big(\theta'_{2s,2s}(z+\nu)+\theta'_{2s,2s}(z-\nu)\big)\ ,
\label{deq}
\ee 
where we have defined the convolution
\EQ{
F\ast G(z)=\frac1{2\pi}\int_{-r}^r dy\,F(z-y)G(y)\ .
}
In the above, the limits on the integral are
\EQ{
r=\infty\qquad(1>|\Delta|)\ ,\qquad \frac\pi2\qquad(\Delta<-1)\ .
}

The Bethe Ansatz equations for the XXZ$_s$ Heisenberg chain have been investigated in the literature \cite{Sogo,Kirillov:1987zz}. The ground state consists of $2s$ strings with no holes, so with a density $\rho_0(z)$ that satisfies
\be
\rho_0(z) +  K_{2s,2s}\ast\rho_0(z)=\frac1{4\pi}\big(\theta'_{2s,2s}(z+\nu)+\theta'_{2s,2s}(z-\nu)\big)\ . 
\label{deq2}
\ee  
This is solved by inverting the kernel which is simple using a Fourier transform
\EQ{
\rho_0(z)=\frac1{4\pi}(I+K_{2s,2s})^{-1}\ast\big(\theta'_{2s,2s}(z+\nu)+\theta'_{2s,2s}(z-\nu)\big)\ .
}
Here, $I$ is the unit integral kernel $\delta(z)$, so $I\ast f=f$.

The simplest physical excitations, the spinons or kinks, correspond to holes in the distribution of $2s$ and their dispersion relation and S-matrix can be extracted by re-writing  \eqref{deq} by swapping over the r\^ole of the strings and the holes on the left-hand side by convoluting with $(I+K_{2s,2s})^{-1}$:
\be
\rho(z) + \rho_h(z) +K^h_{2s,2s}\ast\rho_h (z)\ ,
\label{drh2}
\ee 
where
\EQ{
I+K^h_{2s,2s}=(I+K_{2s,2s})^{-1}\ .
\label{ycc}
}
The hole-hole S-matrix is then determined from
\EQ{
K^h_{2s,2s}(z)=i\frac d{dz}\log S^h_{2s,2s} (z)\ .
}

In order to find out the energy and momentum of a hole with rapidity $z$, one needs to solve \eqref{deq} for the back-reaction of the hole $\rho_h(y)=N^{-1}\delta(y-z)$ on the density $\delta\rho(y)$. Using \eqref{ycc}, one finds  
\EQ{
\delta\rho(y)=-\frac1N\big(\delta(y-z)+K^h_{2s,2s}(y-z)\big)\ .
}
The energy and momentum of the hole excitation is, therefore,
\EQ{
{\cal E}_h(z)=N\int_{-r}^r dy\,\delta\rho(y)\varepsilon(y)\ ,\qquad {\cal P}_h(z)=N\int_{-r}^r dy\,\delta\rho(y)\wp(y)\ ,
}

\vspace{0.2cm}
\noindent{\bf\underline{$1>|\Delta|$}:}

In this regime, our conventions for the Fourier transform is
\EQ{
F(z)=\int_{-\infty}^\infty d\omega\,e^{i\omega z}\tilde F(\omega)\ ,\qquad \tilde F(\omega)=\frac1{2\pi}\int_{-\infty}^\infty dz\,e^{-i\omega z}F(z)\ .
}
The key transform is
\EQ{
\frac1{2\pi}\int_{-\infty}^\infty dz\,e^{-i\omega z}\theta'_j(z)=\frac{\sinh(\omega(\pi-j\gamma)/2)}{\sinh(\omega\pi/2)}\ .
}

In Fourier space, the hole scattering kernel can be expressed via \eqref{ycc}
\EQ{
1+\tilde K^h_{2s,2s}(\omega)=(1+\tilde K_{2s,2s}(\omega))^{-1}=\frac{\sinh(\pi\omega/2)\tanh(\gamma\omega/2)}{2\sinh(s\gamma\omega)\sinh((\pi-2s\gamma)\omega/2)}\ ,
}
and, hence, the hole-hole $S$-matrix takes the form
\EQ{
S^h_{2s,2s}(z)=\exp\Big[2i\int_0^\infty\,\frac{d\omega}\omega\,\sin(\omega z)\Big(1-\frac{\sinh(\pi\omega/2)\tanh(\gamma\omega/2)}{2\sinh(s\gamma\omega)\sinh((\pi-2s\gamma)\omega/2)}\Big)\Big]\ .
\label{sh1}
}
The density of strings in the ground state is
\EQ{
\rho_0(z)=\frac1{4\gamma}\Big(\sech\frac{\pi(z-\nu)}\gamma+\sech\frac{\pi(z+\nu)}\gamma\Big)\ .
}
Finally the energy and momentum of a hole excitation is 
\EQ{
{\cal E}_h(z)&=\frac{\pi}{2L}-\frac1{2L}\arctan\Big(\sinh\frac{\pi(z+\nu)}\gamma\Big)+\frac1{2L}\arctan\Big(\sinh\frac{\pi(z-\nu)}\gamma\Big)\ ,\\
{\cal P}_h(z)&=\frac1{2L}\arctan\Big(\sinh\frac{\pi(z+\nu)}\gamma\Big)+\frac1{2L}\arctan\Big(\sinh\frac{\pi(z-\nu)}\gamma\Big)\ ,
\label{qpp}
}
Note that the energy of the excitation is very different from the conventional Heisenberg XXZ chain.

\vspace{0.2cm}
\noindent{\bf\underline{$\Delta<-1$}:}

In this regime, the Fourier transform is defined via
\EQ{
F(z)=2\sum_{n\in\mathbb Z} e^{2inz}\tilde F(n)\ ,\qquad \tilde F(n)=\frac1{2\pi}\int_{-\pi/2}^{\pi/2} dz\,e^{-2inz}F(z)\ ,
}
and the key transform is
\EQ{
\frac1{2\pi}\int_{-\pi/2}^{\pi/2}dz\,e^{-2inz}\theta'_j(z)=e^{-j\sigma|n|}\ .
}

Now we repeat the analysis of the $1>|\Delta|$ regime. 
The Fourier Transform of the hole scattering kernel is given via
\EQ{
1+\tilde K_h(n)=(1+\tilde K(n))^{-1}=\frac{\tanh(|n|\sigma)}{1-e^{-4|n| s \sigma}} 
}
and, therefore, the hole-hole S-matrix takes the form
\EQ{
S_{hh}(z)=\exp\Big[iz\Big(2-\frac1{2s}\Big)+2i\sum_{n=1}^\infty \frac{\sin(2nz)}n\Big(1-\frac{\tanh(n\sigma)}{1-e^{-4ns\sigma}}\Big)\Big]\ .
\label{sh2}
}
The density of strings in the ground state is
\EQ{
\rho_0(z)&=\frac1{2\pi}+\frac1\pi\sum_{n=1}^\infty\frac{\cos(2n\nu)\cos(2nz)}{\cosh(n\sigma)}\\
&=\frac{  K  }{2\pi^2 }  \Big\{   \text{dn}\Big(\frac{2K (z+\nu)}\pi,\kappa\Big) +  \text{dn}\Big(\frac{2K (z-\nu)}\pi,\kappa\Big)\Big\}\ ,
}
where $K$ and $K'$ are the complete elliptic integrals of modulus $\kappa$ and $\sqrt{1-\kappa^2}$ such that $K'/K=\sigma/\pi$.

The momentum is
\EQ{
{\cal P}_{h}(z)&=-\frac{z}L-\frac1L\sum_{n=1}^\infty \frac{\sin(2nz)\cos(2n\nu)}{n\cosh(n\sigma)}\\ & =  \frac{-1}{2 L}   \Big\{   \text{am}\Big(\frac{2K (z+\nu)}\pi,\kappa\Big) +  \text{am}\Big(\frac{2K (z-\nu)}\pi,\kappa\Big)\Big\}
}
while the energy is
\EQ{
{\cal E}_{h}(z)&=-\frac\nu L-\frac1L\sum_{n=1}^\infty \frac{\cos(2nz)\sin(2n\nu)}{n\cosh(n\sigma)}\\ 
& =   \frac{1}{2 L}   \Big\{ \text{am}\Big(\frac{2K (z+\nu)}\pi,\kappa\Big) - \text{am}\Big(\frac{2K (z-\nu)}\pi,\kappa\Big) \Big\}\ .
}

\subsection{Continuum limits}

The key question for us is whether the spin model has a continuum limit where the lattice spacing $L$ goes to 0 with a fixed correlation length. In order for this limit to exist, the spin model with fixed $L$ must have a gapless phase for some value of the inhomogeneity $\nu$. A coordinated limit can then be taken.

If we look in the $1>|\Delta|$ regime, it is clear from the dispersion relation of the spinon \eqref{qpp}, that the excitation is generally gapped, however, as $\nu\to\infty$, the energy goes to 0 and the system becomes gapless. 
There consequently exists a non-trivial continuum limit where $L\to0$ and $\nu\to\infty$ with 
\EQ{
\frac1L\exp(-\pi\nu/\gamma)=\frac{m}{2}
\label{lcm}
} 
fixed. In this limit, a mass scale $m$ emerges and we obtain a relativistic dispersion relation with $\theta=\pi z/\gamma$ being the relativistic rapidity:
\EQ{
&{\cal E}_h(\theta)=m\cosh \theta\ ,\qquad {\cal P}_h(\theta)=m\sinh\theta\ ,\\
&\text{i.e.}\qquad{\cal E}_h^2-{\cal P}_h^2=m^2\ .
}
This is an example of  {\it dimensional transmutation\/} in QFT, where a mass scale is generated out of a cut off, here $\mu=L^{-1}$, and a dimensionless coupling, here $\nu$. In particular, the beta function of the coupling -- -expressing the way $\nu$ must vary with the cut off to keep the mass scale $m$ fixed -- -is
\EQ{
\mu\frac{d\nu}{d\mu}=\frac\gamma{\pi}\ .
\label{xsi}
}

In the $\Delta<-1$ regime, on the other hand, the fact that the physical quantities are periodic in the rapidity and inhomogeneity means that the system is always gapped and there is no way to take a continuum limit.

\section{S-matrices}
\label{sec:S-matrices}

In this section, we turn to the question of the S-matrix of the excitations of the spin chain. In the last section, we extracted the S-matrix element of holes via the TBA. The analysis, of course, is vastly simplified because it ignores the complications of the full TBA. In particular, the spinons actually come in a spin $\frac12$ multiplet of the deformed $\SU(2)$ symmetry (the affine quantum group $\mathscr U_q(\widehat{\msu(2)})$) of the chain as well as carrying kink quantum numbers that are associated to a different, hidden, quantum group symmetry realized in the IRF/RSOS form.

However, the hole S-matrix element we have extracted in \eqref{sh1} and \eqref{sh2} yields sufficient information that enables us to identify the full S-matrix that manifests the complete symmetry of the theory. The S-matrix is built out of blocks that are associated to the quantum group $\mathscr U_q(\widehat{\msu(2)})$. There are 3 known S-matrix blocks of this type:

\noindent{\bf (i)} The sine-Gordon soliton S-matrix \cite{Zamolodchikov:1978xm}. The particle state is doubly degenerate consisting of the soliton and anti-soliton. The identical particle, e.g.~soliton-soliton, S-matrix $\ket{{\uparrow}(\theta_1){\uparrow}(\theta_2)}\to\ket{{\uparrow}(\theta_2){\uparrow}(\theta_1)}$, $\theta=\theta_1-\theta_2$, can be written
\EQ{
S^I(\theta;\gamma')=\exp\Big\{i\int_0^\infty\frac{dw}w\,\frac{\sin[w\theta]\sinh[\pi w(\gamma'-1)/2]}{\cosh[\pi w/2]\sinh[\pi w\gamma'/2]}\Big\}\ .
\label{g4r}
}
The S-matrix has the affine quantum group $\mathscr U_q(\widehat{\msu(2)})$, with $q=\exp[-i\pi/\gamma']$, symmetry.

\noindent{\bf(ii)} The cyclic sine-Gordon S-matrix \cite{Leclair:2003xj}. This is a kind of variant of the sine-Gordon S-matrix correspond to the quantum group symmetry with $q=\exp[-\pi/\sigma']$ real, describing the scattering of a doublet of particles which is periodic in the rapidity $\theta\to\theta+\pi\sigma'$.\footnote{Actually the S-matrix changes sign under the rapidity shift.} The identical particle S-matrix element is 
\EQ{
S^I_\text{cyclic-SG}(\theta;\sigma')=\exp\Big\{i\theta/\sigma'+i\sum_{n=1}^\infty\frac2n\cdot\frac{\sin[2n\theta/\sigma']}{1+\exp[2\pi n/\sigma']}\Big\}\ .
\label{csg}
}

\noindent{\bf(iii)} The RSOS, or restricted sine-Gordon, S-matrix \cite{LeClair:1989wy,Bernard:1990ys,Bernard:1990cw}. The S-matrix is related to the sine-Gordon S-matrix by performing a vertex-to-IRF transformation and then a restriction from the SOS to RSOS form of the associated R-matrix. This can also be viewed as the restriction to the special representations of $\mathscr U_q(\widehat{\msu(2)})$ which exist for the deformation paramater $q$ a root of unity, $q=\exp[-i\pi/(k+2)]$. The particle is then realized as a doublet of kinks $K_{ab}(\theta)$, where $b=a\pm\frac12$ and the ``vacua" $a,b\in\{0,\frac12,1,\frac32,\ldots,\frac k2\}$. The identical particle, i.e
\EQ{
\ket{K_{a+1,a+\frac12}(\theta_1)K_{a+\frac12,a}(\theta_2)}\to\ket{K_{a+1,a+\frac12}(\theta_2)K_{a+\frac12,a}(\theta_1)}\ ,
} 
S-matrix element is
\EQ{
S^I_{\text{RSOS}}(\theta;k)=\exp\Big\{i\int_0^\infty\frac{d\omega}\omega\,\frac{\sin[\omega\theta]\sinh[\pi\omega(k+1)/2]}{\cosh[\pi\omega/2]\sinh[\pi\omega(k+2)/2]}\Big\}\ .
\label{sro}
}

In order to gain some intuition on the form of the S-matrix, we can examine the S-matrix of the higher spin XXX Heisenberg chain. This is the $\Delta=1$ point of the XXZ model. The excitations of this model transform in a doublet of the $\SU(2)$ (so independent of the spin $s$ of the chain) but they also carry kink quantum numbers that depend on $s$. The S-matrix has the characteristic product form \cite{Reshetikhin:1990jn}
\EQ{
S(\theta) = S_{\SU(2)}(\theta) \otimes S_\text{RSOS}(\theta;2s)\ ,
}
where the first factor here is the $\SU(2)$ invariant S-matrix block (actually Yangian invariant). This is the sine-Gordon S-matrix in the limit $\gamma'\to\infty$. So we can expect the S-matrix of the higher spin XXZ chain to have this characteristic product form, where the first factor manifests the symmetry of the chain, in this case the affine quantum group $\mathscr U_q(\widehat{\msu(2)})$. The second factor describes the hidden kink structure of the spinons.

The question is, what S-matrix elements should be compared with the hole-hole S-matrix element we have calculated in the spin chain? This turns out to be quite a subtle issue. The simple answer is that it corresponds to the scattering of two identical states of the full S-matrix. However, the complete answer is found by constructing TBA equations for the putative S-matrix and then comparing with the TBA system of the spin chain around the anti-ferromagnetic vacuum state. The hole excitation is then identified with highest weight states of the full TBA, but the effective S-matrix kernel is not quite the S-matrix of the highest weight state due to the associated magnon system. It turns out that the level $k$ of the RSOS part has a characteristic shift: $k\to k-2$ (see appendix \ref{a3} for an explanation). So the hole-hole S-matrix should be compared with the identical particle S-matrix but with a shift $k\to k-2$ in the RSOS factor.

\subsection{S-matrix of the $1>|\Delta|$ (UV safe) regime}

In this regime, one immediately identifies the hole-hole scattering kernel with the identical kink element of the product S-matrix (after the shift $k\to k-2$, as explained in appendix \ref{a3})
\be
  S = S_\text{SG}(\theta; \gamma' ) \otimes S_\text{RSOS}(\theta;k)\ ,
  \label{rrp}
  \ee
where
\be
2 s = k \ , \qquad  \gamma' =\frac\pi\gamma-k  \ .
\label{yup}
\ee
Note that we now have the exact relation between $\gamma'$ and $\gamma$ for which \eqref{rgg} is the $k\to0$ limit. This is exactly the S-matrix of the fractional supersymmetric sine-Gordon theory \cite{Bernard:1990ti} that we proposed in \cite{Appadu:2017bnv} to describe the S-matrix of the UV safe XXZ lambda model. We now see that the light-cone lattice spin chain provides a bottom up approach to quantize this theory.

There is an important overall consistency condition that arises from this.
Recall from the continuum limit we found that the inhomegenity parameter must vary with scale    leading to the beta function \eqref{xsi}
\EQ{ \mu\frac{d\nu}{d\mu} = \frac\gamma\pi\ . } 
This result should be matched to RG flow of the  XXZ lambda model near the UV (where $\lambda \to \lambda_\star$ and $\xi$ is small).  Now from  \eqref{inh}, the inhomogeneity parameter behaves in UV as
\EQ{
\nu \approx \frac12\log\frac{2\lambda_*}{\xi(1-\lambda_*)}=\frac12\log\frac k{\gamma'\xi}\ ,
}
and hence
\EQ{
\mu\frac{d\nu}{d\mu}\approx -\frac1{2\xi}\cdot\mu\frac{d\xi}{d\mu}  \ . 
}
Now we make use of the beta function for $\xi$  eq.~\eqref{rg2}  in the UV  where $\xi$ is small to find 
\EQ{
\mu\frac{d\nu}{d\mu}\approx \frac1{\gamma'+k} \ . 
}
The two above expressions for the beta function match precisely upon the use of the identification of parameters in eq.~\eqref{yup}. 

Since the S-matrix \eqref{rrp} includes the sine-Gordon kink S-matrix block there are bound states, the breathers, in the regime when $\gamma'<1$. This corresponds to  
\EQ{
\frac\pi\gamma-2s<1\ ,
}
in the spin chain. The breathers have a mass spectrum
\EQ{
M_n=2m\sin\frac{n\pi\gamma'}2\ ,\qquad n=1,2,\ldots\, <\frac1{\gamma'}\ .
}
It follows that the kinks are not the lightest states in the spectrum when $2\sin(\pi\gamma'/2)<1$. 

In the spin $\frac12$ XXZ chain, in the limit $\gamma\to\pi$ the XXZ chain is approaching the ferromagnetic XXX point $\Delta=1$. In this limit, the breathers are identified with the spin waves of the ferromagnetic XXX chain.
We will discuss the breathers more fully in section \ref{s6}.

\subsection{S-matrix of the $\Delta<-1$ (cyclic) regime: the missing block}\label{s5.3}

Now we turn to the $\Delta<-1$ regime. In this case, the hole-hole S-matrix element \eqref{sh2}, like the whole TBA, is periodic in the rapidity $z$.  This periodicity should be reflected in the complete S-matrix.  We anticipate that the S-matrix should consist of the product of two blocks, the first of which should be the cyclic SG S-matrix block reflecting the  quantum group symmetry of the chain and the second should encode the kink structure of states.   But the conventional RSOS S-matrix is not, as is clear in \eqref{sro}, periodic in rapidity. This hints that there should be a new S-matrix block of the RSOS type which is also periodic in rapidity.  In this section we shall first detail the construction of this new block with which we shall then  describe  the S-matrix of the $\Delta<-1$ regime.

The conventional {\it trigonometric\/} RSOS kink S-matrix is built out of solution to the Yang-Baxter Equation, or more precisely the {\it star-triangle relation\/}, that play the role of Boltzmann weights in an Interaction Round a Face (IFR) statistical model.    These Boltzmann weights have an elliptic generalization, e.g.~see \cite{Jimbo:1987ra},  that will allow us to construct the new S-matrix with the required periodicity. It is also noteworthy that the elliptic IRF models are associated to any classical Lie group, although for present purposes we need only concentrate on $\SU(2)$, or more precisely the affine quantum group $\mathscr U_q(\widehat{\msu(2)})$.

In the IRF S-matrix, the states are kinks $K_{ab}(\theta)$ and states are labelled by the vacua $a,b$ on either side.  The vacua (the local heights of the statistical model) are associated to representations of $\mathscr U_q(\msu(2))$ so to spins $a,b,\ldots\in\{0,\frac12,1,\frac32,\ldots\}$. When $q$, the quantum group parameter, is a root of unity, 
\EQ{
q=\exp\big[-i\pi/(k+2)\big]\ ,
}
there is a restricted model, where the spins are restricted to lie in the set of integrable representations of level $\leq k$, so $a,b,\ldots\in\{0,\frac12,1,\dots,\frac k2\}$. A basis of states in the Hilbert space with $N$ kinks is labelled by a sequence $\{a_{N+1},a_N,\ldots,a_1\}$, which has the interpretation of a {\it fusion path\/}, so the spin $a_{j+1}$ representation must appear in the tensor product of the $a_j$ representation with the spin $\frac12$ representation (truncated by the level restriction). This means that there is an adajency condition $a_{j+1}=a_j\pm\frac12$. 

The analogue of the $R$-matrix, is an intertwiner $W$ between  2-kink states \cite{Jimbo:1987ra}:
\EQ{
\ket{K_{ab}(\theta_1)K_{bc}(\theta_2)}\longrightarrow \sum_d W\left.\left.\left(\hspace{-0.2cm}\raisebox{-14.5pt}{\begin{tikzpicture}[scale=0.32]
\node at (0,1.1) {$d$};
\node at (0,-1.1) {$b$};
\node at (1.2,0) {$c$};
\node at (-1.2,0) {$a$};
\end{tikzpicture}}\hspace{-0.15cm}\right.\right| u\right)\ket{K_{ad}(\theta_2)K_{dc}(\theta_1)}\ ,
}
where $u=\theta/(i\pi)$ and $\theta=\theta_1-\theta_2$. These intertwiners satisfy the star triangle relation \cite{Jimbo:1987ra}:
\EQ{
 &\sum_{x} \WW{a}{b}{c}{x}{u} \WW{x}{c}{d}{e}{u+v} \WW{a}{x}{e}{f}{v}  \\
& \quad =  \sum_{x} \WW{b}{c}{d}{x}{v} \WW{a}{b}{x}{f}{u+v} \WW{f}{x}{d}{e}{u} \ .
}
The solution of the star triangle relation $W(u)$ is the raw fodder from which we will fashion an S-matrix for kinks states. There are 3 basic types of non-vanishing elements that take the form
\EQ{
W\left.\left.\left(\hspace{-0.2cm}\raisebox{-21.5pt}{\begin{tikzpicture}[scale=0.5]
\node at (0,1) {\footnotesize $a\pm\frac12$};
\node at (0,-1) {\footnotesize $a\pm\frac12$};
\node at (1.3,0) {\footnotesize $a$};
\node at (-1.1,0) {\footnotesize $a\pm1$};
\end{tikzpicture}}\hspace{-0.15cm}\right.\right|u\right)&=\frac{[1-u]}{[1]}\ ,\\[5pt]
W\left.\left.\left(\hspace{-0.2cm}\raisebox{-21.5pt}{\begin{tikzpicture}[scale=0.5]
\node at (0,1) {\footnotesize $a\pm\frac12$};
\node at (0,-1) {\footnotesize $a\pm\frac12$};
\node at (1.5,0) {\footnotesize $a$};
\node at (-1.5,0) {\footnotesize $a$};
\end{tikzpicture}}\hspace{-0.15cm}\right.\right|u\right)&=\frac{[\pm(2a+1)+u]}{[\pm(2a+1)]}\ ,\\[5pt]
W\left.\left.\left(\hspace{-0.2cm}\raisebox{-21.5pt}{\begin{tikzpicture}[scale=0.5]
\node at (0,1) {\footnotesize $a\mp\frac12$};
\node at (0,-1) {\footnotesize $a\pm\frac12$};
\node at (1.5,0) {\footnotesize $a$};
\node at (-1.5,0) {\footnotesize $a$};
\end{tikzpicture}}\hspace{-0.15cm}\right.\right|u\right)&=-\frac{[u]}{[1]}\cdot\frac{\sqrt{[2a+2][2a]}}{[2a+1]}\ ,
\label{wwd}
}
where we have defined
\EQ{
[u]=\theta_1\Big(\frac {i\pi u}{\sigma'};\mathfrak q\Big)\ ,
}
in terms of the Jacobi theta function  
\EQ{
\theta_1(x;\frak q)= 2 \mathfrak q^{1/4 }\sin x\prod_{n=1}^\infty(1-2\mathfrak q^{2n}\cos2x+\mathfrak q^{4n})(1-\mathfrak q^{2n})\ .
\label{tff}
} 
Here, the elliptic nome $\frak q$ is not to be confused with a quantum group parameter and the parameter $\sigma^\prime$ is allowed to be an arbitrary real number that we shall shortly associate with the RG invariant of the $XXZ$ lambda theory in the cyclic regime $\gamma^\prime = i \sigma^\prime$.

The $W$ map satisfies some identities that are important for the S-matrix that we going to build \cite{Jimbo:1987ra}:

\noindent
{\bf(i)} the {\it initial condition\/}
\EQ{
W\left.\left.\left(\hspace{-0.2cm}\raisebox{-14.5pt}{\begin{tikzpicture}[scale=0.32]
\node at (0,1) {$d$};
\node at (0,-1) {$b$};
\node at (1.2,0) {$c$};
\node at (-1.2,0) {$a$};
\end{tikzpicture}}\hspace{-0.15cm}\right.\right|0\right)=\delta_{bd}\ ;
}

\noindent
{\bf(ii)} {\it rotational symmetry\/}
\EQ{
W\left.\left.\left(\hspace{-0.2cm}\raisebox{-14.5pt}{\begin{tikzpicture}[scale=0.32]
\node at (0,1) {$d$};
\node at (0,-1) {$b$};
\node at (1.2,0) {$c$};
\node at (-1.2,0) {$a$};
\end{tikzpicture}}\hspace{-0.15cm}\right.\right|1-u\right)=\sqrt{\frac{[2b+1][2d+1]}{[2a+1][2c+1]}}W\left.\left.\left(\hspace{-0.2cm}\raisebox{-13.5pt}{\begin{tikzpicture}[scale=0.32]
\node at (0,1) {$c$};
\node at (0,-1.2) {$a$};
\node at (1.2,0) {$b$};
\node at (-1.2,0) {$d$};
\end{tikzpicture}}\hspace{-0.15cm}\right.\right|u\right)\ ;
}

\noindent
{\bf(iii)} {\it inversion relation\/}
\EQ{
\sum_d W\left.\left.\left(\hspace{-0.2cm}\raisebox{-14.5pt}{\begin{tikzpicture}[scale=0.32]
\node at (0,1) {$d$};
\node at (0,-1) {$b$};
\node at (1.2,0) {$c$};
\node at (-1.2,0) {$a$};
\end{tikzpicture}}\hspace{-0.15cm}\right.\right|u\right)
W\left.\left.\left(\hspace{-0.2cm}\raisebox{-14.5pt}{\begin{tikzpicture}[scale=0.32]
\node at (0,1) {$e$};
\node at (0,-1) {$d$};
\node at (1.2,0) {$c$};
\node at (-1.2,0) {$a$};
\end{tikzpicture}}\hspace{-0.15cm}\right.\right|-u\right)=\frac{[1-u][1+u]}{[1]^2}\delta_{be}\ .
}
The alert reader will recognize that the rotational symmetry and inversion relation as proto-identities for crossing symmetry and braiding unitarity, respectively, of the S-matrix we will construct.

We have defined the $u$ dependence so that $W(u)$ has a manifest periodicity that corresponds to the rapidity shifts $\theta\to\theta+\pi\sigma'$, as we require of our S-matrix.\footnote{Under $\theta\to\theta+\pi\sigma'$ we have $W(u)\to  - W(u)$.} However, having fixed the scaling like this we would seem to have violated the definition needed to have a restricted model, i.e.~having local heights in the finite set $\{0,\frac12,1,\ldots,\frac k2\}$.   The restricted model  requires that $[0]=[k+2]=0$ so that $W(u)$ cannot propagate a kink state with admissible local heights $\ket{K_{ab}(\theta_1)K_{bc}(\theta_2)}$ with $a,b,c\in\{0,\frac12,1,\ldots,\frac k2\}$ into one with an inadmissible local height $\ket{K_{ad}(\theta_2)K_{dc}(\theta_1)}$ with $d\not\in\{0,\frac12,1,\ldots,\frac k2\}$, i.e.~$d=0$ or $\frac k2+1$.

The conventional restricted sine-Gordon, or RSOS, S-matrix \cite{LeClair:1989wy}, is built from the same solution of the star-triangle relation with $\sigma'=-i(k+2)$   in the trigonometric limit where the nome $\frak q\to0$, so that $[u]\to-2\frak q^{1/4}\sin(\pi u/(k+2))$.  In that case it is immediate to see that   precisely  $[0]=[k+2]=0$ so the restriction to the RSOS model is consistent.  Here in contrast we shall exploit the fact that the Jacobi theta function has a large set of zeros at $\theta_1(\p(m +n\tau);\mathfrak q)=0$, $m,n\in\mathbb Z$, where the nome is $\mathfrak q=\exp[i\pi\tau]$.  This can be used to  define a different restricted model.   In particular if we choose   
\EQ{
\tau=\frac{i(k+2)}{\sigma'}\ ,\qquad \text{i.e.}\quad \mathfrak q=\exp\big[-\pi(k+2)/\sigma'\big]\  ,
}
then we ensure simultaneously the periodicity $[u + i \sigma^\prime ] = - [u]$ and the consistency of the restricted model $[0]=[k+2]=0$. 

There should be the sense that if $\sigma'\to\infty$   then we get the usual restricted sine-Gordon theory with its trigonometric RSOS S-matrix. In order to take that limit, we have to take an ``S-duality" transformation of the theta functions,
\EQ{
\theta_1\big(x;\mathfrak q\big)=- (i\tau)^{-1/2}\exp\big[-i  x^2/(\pi \tau) \big]\theta_1\big(x/\tau;\hat{\mathfrak q}\big)\ ,
}
where $\hat{\mathfrak q}=\exp[-i\pi/\tau]$. Now the limit $\sigma\to\infty$ corresponds to $|\tau|\to0$ and so $\hat{\mathfrak q}\to0$. Then, using \eqref{tff} (up to some constant factors that cancel in ratios) the limit corresponds to 
\EQ{
[u]\longrightarrow\sin\Big(\frac{\pi u}{k+2}\Big)\ ,
}
which is the usual trigonometric limit yielding the RSOS S-matrix.

In order to make a consistent S-matrix, what we call the {\it elliptic RSOS S-matrix\/},
\EQ{
S_\text{ell-RSOS}(\theta;\sigma',k)=v(\theta)W\left.\left.\left(\hspace{-0.2cm}\raisebox{-14.5pt}{\begin{tikzpicture}[scale=0.32]
\node at (0,1) {$d$};
\node at (0,-1) {$b$};
\node at (1.2,0) {$c$};
\node at (-1.2,0) {$a$};
\end{tikzpicture}}\hspace{-0.15cm}\right.\right|u(\theta)\right)\ ,
}
we have to construct a suitable scalar factor $v(\theta)$ in order that the S-matrix is unitary and crossing symmetric. The scalar factor must satisfy
\EQ{
v(\theta)=v(i\pi-\theta)\ ,\qquad v(\theta)v(-\theta)=\frac{\theta_1\big(i\pi/\sigma';\mathfrak q\big)^2}{\theta_1\big((i\pi-\theta)/\sigma';\mathfrak q\big)\theta_1\big((i\pi+\theta)/\sigma';\mathfrak q\big)}\ .
}
In order to solve these conditions, we apply the usual minimality assumption to find the solution
\EQ{
v(\theta;\sigma',k)=\exp\Big\{\sum_{n=1}^\infty\frac2n\cdot\frac{\cosh(\pi nk/(2\sigma'))\sin(n(i\pi-\theta)/(2\sigma'))
\sin(n\theta/(2\sigma'))}{\cosh(\pi n/(2\sigma'))\sinh(\pi n(k+2)/(2\sigma'))}\Big\}\ .
}

The identical particle S-matrix element, i.e.~for the process
\EQ{
\ket{K_{a+1,a+\frac12}(\theta_1)K_{a+\frac12,a}(\theta_2)}\to\ket{K_{a+1,a+\frac12}(\theta_2)K_{a+\frac12,a}(\theta_1)}\ ,
} 
is then
\EQ{
S^I_{\text{ell-RSOS}}(\theta;\sigma',k)=\exp\Big\{i\theta/\sigma'+i\sum_{n=1}^\infty\frac1n\cdot\frac{\sin[2n\theta/\sigma']\sinh[\pi n(k+1)/\sigma']}{\cosh[\pi n/\sigma']\sinh[\pi n(k+2)/\sigma']}\Big\}\ .
\label{szz}
}

The elliptic RSOS S-matrix is related to other S-matrix blocks in various ways. Firstly, in the limit $\sigma'\to\infty$, the sum in to \eqref{szz} turns into an integral, $\sum_nn^{-1}\to \int dw\,w^{-1}$, with $w=n/\sigma'$. The S-matrix becomes the conventional RSOS S-matrix. In particular, \eqref{szz} becomes \eqref{sro}:
\EQ{
S_{\text{ell-RSOS}}(\theta;\sigma',k)\overset{\sigma'\to\infty}{\xrightarrow{\hspace*{1.5cm}}}S_\text{RSOS}(\theta;k)\ .
}

The cyclic sine-Gordon S-matrix, can be obtained in the limit $k\to\infty$, followed by an IRF-to-vertex transformation. In particular. \eqref{szz} becomes \eqref{csg}:
\EQ{
S_{\text{ell-RSOS}}(\theta;\sigma',k)\overset{\substack{k\to\infty\\[2pt] \text{IRF-to-vertex}}}{\xrightarrow{\hspace*{3cm}}}S_\text{cyclic-SG}(\theta;\sigma')\ .
}

Finally, we turn to the S-matrix of spin $s$ XXZ spin chain in the $\Delta<-1$ regime and the cyclic XXZ lambda model. First of all, by looking at the hole-hole S-matrix of the spin chain and the S-matrix block we have identified and constructed, we find that the spin chain S-matrix is precisely the product
\be
  S = S_\text{cyclic-SG}(\theta; \sigma' ) \otimes S_\text{ell-RSOS}(\theta;\sigma',k)\ ,
  \ee
where
\be
2 s = k \ , \qquad  \sigma' =\frac\pi\sigma\ ,
\label{yup2}
\ee
and where the rapidities are related via
\EQ{
z=\frac\theta{\sigma'}\ .
}
The S-matrix is then periodic in rapidity $\theta\to\theta+\pi\sigma'$.

It is clear that this is not the same as the proposed S-matrix of the cyclic XXZ lambda model in \eqref{sm9}; the RSOS factors differ: the spin chain involves the elliptic R-matrix while the lambda model the conventional one. If we change the proposal \eqref{sm9} to involve the elliptic RSOS block instead, then the lambda model S-matrix does not yield the sigma model S-matrix in the limit $k\to\infty$ (and after an IRF-to-vertex transformation). So there is a fundamental problem in pulling together the S-matrix of the QFT based on symmetries and the underlying spin chain. 

The mismatch between the S-matrix suggests that the XXZ lambda model in the cyclic regime cannot be regularized as a spin chain and a continuum obtained by taking a scaling limit.
This is not completely definitive, but adds weight to the idea that theories with a cyclic RG do not actually have a continuum limit and only exist as effective theories. Previous evidence involved calculating finite-size effects in a theory with a cyclic RG  \cite{LeClair:2004ps}. This analysis suggested that there was a singularity in the deep UV. The implications of this result for the present interest in integrable deformations of sigma and lambda models is wide ranging because it would imply that Yang-Baxter type deformations do not define continuum, integrable quantum field theories.

\section{Breathers and the $\SU(2)/\U(1)$ limit}\label{s6}

In this section, we consider the breather states in the spectrum of the XXZ lambda model and spin chain that exist in the spectrum of the UV safe regime when $\gamma'<1$. We are particularly interested in the limit $\gamma'\to0$, corresponding to $\lambda\to1$. In this limit, the XXZ lambda model has an emergent $\U(1)$ gauge symmetry and becomes the symmetric space $\SU(2)/\U(1)$ lambda model \cite{Sfetsos:2013wia,Hollowood:2014rla}. The interesting question is how the emergent gauge symmetry is reflected in the spectrum of the theory and it is to this question that we now turn. 

As we shall see the situation is very closely related to the way that the states and S-matrix of the $\text{O}(3)$ sigma model are related to the states and S-matrix of the anisotropic XXZ $\SU(2)$ principal chiral model \cite{Karowski:1978iv,Wiegmann:1985jt}. This was viewed as a simple example of confinement where the states of the $\text{O}(3)$ model are the analogues of the mesons formed as bound states of the excitations of the XXZ model, the analogues of the quarks. In fact, we recover this scenario in the limit $k\to\infty$.

The pole structure of the S matrix comes from the sine-Gordon block in \eqref{rrp} but the states also carry RSOS kink quantum numbers. The kinks are built on a Hilbert space which is spanned by vectors associated to {\it fusion paths\/}, so strings of $\SU(2)$ spins, $[a_N,a_{N-1},\ldots,a_1]$, $a_i\in\frac12\mathbb Z\geq0$, with the adjacency condition $a_{p+1}=a_p\pm\frac12$ and the height restriction $a_i\leq\frac k2$. The kinks transform in the spin $\frac12$ representation and have the form $K_{ab}(\theta)$ with paths of length $\frac12$: $a=b\pm\frac12$. The RSOS S-matrix block has the spectral decomposition
\EQ{
S_\text{RSOS}(\theta;k)=f(\theta;k)\Big\{\sinh\Big(\frac{\theta-i\pi}{k+2}\Big)\mathbb P^{[1]}+\sinh\Big(\frac{\theta+i\pi}{k+2}\Big)\mathbb P^{[0]}\Big\}\ .
\label{rnn}
}
where $f(\theta;k)$ is a scalar factor that ensures unitarity and crossing. The two terms in the above, correspond to the triplet and singlet, representations. The poles in the soliton anti-soliton scattering of the sine-Gordon piece corresponding to the breather $B_n$ occur at
\EQ{
\theta_n=i\pi(1-n\gamma')\ ,\qquad n=1,2,\ldots<\frac1{\gamma'}\ .
}
Each breather must therefore include states in the RSOS factor in the reducible representation $[\frac12]^2=[1]\oplus[0]$. However, since the sine-Gordon factor implies that the breather $B^{(m+n)}$ can be formed as a bound state of $B^{(m)}$ and $B^{(n)}$ means that for consistency the breather $B^{(n)}$ must carry RSOS quantum numbers in the reducible representation $[\frac12]^{2n}$. In terms of the RSOS Hilbert space, the breather state $B^{(n)}_{a_n\cdots a_1}$ corresponds to paths $(a_n\cdots a_1)$ of length $n$ (including paths that back track on themselves) and subject to the RSOS restriction $a_i\in\{0,\frac12,1,\ldots,\frac k2\}$. For example, $B^{(1)}$ has 4 states $(a+1,a+\frac12,a)$, $(a,a+\frac12,a)$, $(a,a-\frac12,a)$ and $(a-1,a-\frac12,a)$.

The S-matrix element of the breather can be found by applying the bootstrap equations \cite{Karowski:1978iv}. Here, we will content ourselves with writing down the S-matrix for the scattering of two the basic breathers $B^{(1)}$ states. The breather $B^{(1)}$ forms at the pole of the sine-Gordon soliton-anti-soliton part of the S-matrix at $\theta=i\pi(1-\gamma')$. So the breather is a soliton/anti-soliton bound state but with RSOS kink quantum numbers. According to the bootstrap equations, which we review in appendix \ref{a2}, the S-matrix element for scattering of the two breathers is equal to the scattering of the four constituent (anti-)solitons but with appropriate projectors. So altogether there are four solitons with rapidities $\vartheta_i$, $i=1,2,3,4$. The first/second two make up the first/second breather and so
\EQ{
\vartheta_1=\theta_1+i\pi(1-\gamma')/2\ ,\qquad \vartheta_2=\theta_1-i\pi(1-\gamma')/2\ ,\\
\vartheta_3=\theta_2+i\pi(1-\gamma')/2\ ,\qquad \vartheta_4=\theta_2-i\pi(1-\gamma')/2\ ,
}
where the two breathers have rapidities $\theta_1$ and $\theta_2$. The S-matrix for the scattering of the two breathers is, schematically,
\EQ{
S_{B^{(1)}B^{(1)}}(\theta)=S^\text{SG}_{11}\otimes R_{12}^{-1/2}R_{34}^{-1/2}S_{23}S_{13}S_{24}S_{14}R_{12}^{1/2}R_{34}^{1/2}\ .
}
In the above, $S^\text{SG}_{11}$ is the sine-Gordon breather-breather S-matrix  and $S_{ij}=S_\text{RSOS}(\vartheta_i-\vartheta_j)$ is the RSOS S-matrix for the scattering of the kink consistuents. The projector factor ensures that the bound state is formed with the correct couplings and is precisely the RSOS S-matrix at the breather pole:
\EQ{
R=S_\text{RSOS}(i\pi(1-\gamma'))=r_{[1]}\mathbb P^{[1]}+r_{[0]}\mathbb P^{[0]}\ .
}

We can now see exactly what happens as we approach the limit $\gamma'\to0$ where we expect an emergent gauge symmetry. In this limit, the solitons (the analogues of the quarks in the confinement scenario) become finitely heavy and decouple and the higher breathers become unbound leaving only the lightest breather state $B^{(1)}$ (the meson), transforming as a quartet of kinks. In addition, as $\gamma'\to0$, it follows from \eqref{rnn} that $r_{[1]}\to0$ which signals fact that the triplet and singlet states mutually decouple. In order to see this, in the limit one has $R_{12}=S_{12}\propto\mathbb P^{[0]}$ and $R_{34}=S_{34}\propto\mathbb P^{[0]}$. Then one can move the projector through the S-matrix from right to left using the Yang-Baxter equation, until it hits the projectors on the left. It follows that, in the limit, we have
\EQ{
R_{12}^{-1/2}&R_{34}^{-1/2}S_{23}S_{13}S_{24}S_{14}R_{12}^{1/2}R_{34}^{1/2}\\
&\longrightarrow \mathbb P_{12}^{[0]}\mathbb P_{34}^{[0]}S_{23}S_{13}S_{24}S_{14}\mathbb P_{12}^{[0]}\mathbb P_{34}^{[0]}+\mathbb P_{12}^{[1]}\mathbb P_{34}^{[1]}S_{23}S_{13}S_{24}S_{14}\mathbb P_{12}^{[1]}\mathbb P_{34}^{[1]}\ .
}
It is natural to identify this decoupling as a symptom of the emergent gauge symmetry. The remaining physical state is then the triplet component $B^{(1)}_{[1]}$ and, taking into account that $S_{11}^\text{SG}\to1$, with an S-matrix
\EQ{
S_{B^{(1)}_{[1]}B^{(1)}_{[1]}}(\theta)=\mathbb P^{[1]}_{12}\mathbb P^{[1]}_{34}S_{23}(\theta-i\pi)S_{13}(\theta)S_{24}(\theta)S_{14}(\theta+i\pi)\mathbb P^{[1]}_{12}\mathbb P^{[1]}_{34}\ ,
}
where we have restored the rapidity dependence explicitly.

If we now take the $k\to\infty$ limit, and an IRF-to-vertex transformation, the S-matrix becomes precisely that of the $\text{O}(3)$  (or $\SU(2)/\U(1)$) sigma model, whose states transform in the vector (triplet) representation \cite{Zamolodchikov:1978xm}. This is precisely the original picture established in the old literature \cite{Karowski:1978iv,Wiegmann:1985jt}.

\section{Finite Density TBA}
\label{sec:TBA}
While the symmetries of an integrable QFT are a powerful tool to constrain the S-matrix, whenever possible one should subject the proposed S-matrix to non-trivial tests. In this section we will consider a version of the TBA at zero temperature but with a large chemical potential.\footnote{Calculations of this type can be found in \cite{Hasenfratz:1990zz,Forgacs:1991rs,Forgacs:1991ru,Evans:1995dn,Evans:1994sy,Evans:1994sv,Evans:1994hi} and in particular \cite{Zamolodchikov:1995xk} whose approach we will follow closely.} The chemical potential $h$ corresponds to coupling the theory to a globally conserved $\U(1)$ charge through a modification of the Hamiltonian $H\to H-hQ$. For large $h$, particles with positive charge condense in the vacuum in a way that can be calculated directly from the S-matrix. On the other hand, large $h$ drives the running couplings towards the UV fixed point, and if $k$ is large perturbation theory can be applied. The advantage of this scenario is that it will allow us to essentially extract the exact 1-loop beta function from the S-matrix.

\subsection{The perturbative treatment}
 
The Lagrangian of eq.~\eqref{eq:act} has a $\U(1)$ that acts vectorially $\CF\to h \CF h^{-1}$. The  coupling to the charge can be achieved    by gauging this symmetry with  a constant background temporal component for the   gauge field.  This is implemented by replacing the WZW model with the appropriate gauged WZW and in the deforming term replacing partial derivative with
\EQ{
\partial_\pm\CF\to \partial_\pm \CF+\frac{ih}2[\sigma_3,\CF]\ .
}
This chemical potential induces a potential for the field variables $\CF$.  To be concrete let us parametrize the group element as
\EQ{
\CF=\MAT{-\sin\theta-i\cos\theta\sin\phi & i\cos\theta\cos\phi e^{i\psi}\\   i\cos\theta\cos\phi e^{-i\psi} &
-\sin\theta+i\cos\theta\sin\phi}\ ,
}
for which the induced potential takes the form
\EQ{
V&=\frac{kh^2}{\pi}(1-\xi^2)(\lambda-1)\cos^2\theta\cos^2\phi\Big((1-\xi)^2(1-\lambda)+\\ &\qquad4\xi(1-\lambda)\cos^2\theta+2(1+\xi)(\lambda-\xi)\cos^2\theta\cos^2\phi\Big)^{-1}\ .
}
This potential has a minimum is $\theta=0$, $\phi=0$ and $\psi$ arbitrary, 
\EQ{
\CF_0=\MAT{0 & ie^{i\psi}\\ ie^{-i\psi}&0}\ .
}
It is important for our analysis that the ground state has a tree-level energy density 
\EQ{
V_0=\frac{kh^2}\pi\cdot\frac{\lambda-1}{\lambda+1}\ .
\label{fre}
}
Let us remark that this depends on the coupling $\lambda$, but not $\xi$. 

Now we turn to the quantum theory. The tree-level potential \eqref{fre}, along with the one-loop beta function \eqref{rg2} or \eqref{rg3} allows us to calculate the ground state energy density shift $\delta{\cal E}(h)={\cal E}(h)-{\cal E}(0)$ in the following non-trivial scaling limit. In order to describe it, consider the RG equation \eqref{rg3} with the RG scale identified with the chemical potential because that is the relevant mass scale in our problem. The scaling limit involves:
\EQ{
&k\to\infty\ ,\qquad\gamma'\to\infty\ ,\qquad h\to\infty\\
&k^{-1}\log h=\text{fixed}\ ,\qquad k/\gamma'=\text{fixed}\ .
\label{scl}
}
The limit $k\to\infty$ suppress higher loops, while keeping $\gamma'/k$ fixed means that the UV limit
\EQ{
\lambda_*=\frac{k}{2\gamma'+k }
}
is fixed. So the scaling limit is in the far UV. In this limit, the ground state energy density shift is equal to the tree level \eqref{fre} with the one-loop running coupling $\lambda(h)$ inserted. 

We can write the result in a compact way by defining a new coupling $f$ via
\EQ{
f=\frac{\lambda_*+1}{\lambda_*-1}\cdot\frac{\lambda-1}{\lambda+1} \ , 
}
such that  $f \to 1$ in the UV  where $\lambda \to \lambda_*$ .  We also introduce the parameters,
\EQ{
\kappa=\frac{\gamma'-k}{\gamma'+k}=\frac{3\lambda_*-1}{\lambda_*+1}\ ,\qquad \delta=\frac2{\gamma'+k}=\frac{4\lambda_*}{k(\lambda_*+1)}\ .
}
and the renormalization group parameter
\EQ{
Q=\Big(\frac\Lambda h\Big)^\delta\ ,
}
where $\Lambda$ has unit mass dimension and is the ``lambda parameter" (i.e. the dynamically generated scale) of our renormalization group scheme. The reason for the definition of $\delta$ is that the beta function of  eq.~\eqref{rg2}, which becomes exact in the scaling limit, is solved with a power series
\be
\lambda(h) = \lambda_\star + \sum_{n=1}^\infty \lambda_n Q^{2n}  \ . 
\ee

The beta function for $f$ in the scaling limit can then be written using eq.~\eqref{rg2}
\EQ{
Q\frac{\partial f}{\partial Q}=-\frac{(1-f^2)(1-\kappa^2f^2)}{1-\kappa^2}\ ,
\label{tuy}
}
while the ground state energy shift takes the form
\EQ{
\delta{\cal E}(h)=-\frac{h^2\gamma' k}{\pi(\gamma'+k)}f\ ,
\label{rff}
}
where $f$ solves \eqref{tuy}.  One can integrate the beta function to find $f$ implicitly in terms of $Q$
\EQ{
\frac{1-f}{1+f}\cdot\Big(\frac{1+\kappa f}{1-\kappa f}\Big)^\kappa=Q^2\ .
}
Thus the solution takes the form of an infinite series
\EQ{
f=1+\sum_{n=1}^\infty f_nQ^{2n}\ .
}
More precisely we have 
\be
f_n = \frac{(1- \kappa)^{n (\kappa -1) +1}}{ (1+ \kappa)^{n (\kappa +1) - 1}} g_n(\kappa)
\ee
where $g_n$ is a polynomial of degree $2n-1$. We can now interpret $\Lambda$ as the  integration constant implicit in $Q$ and having fixed it as above we have  defined our renormalization group scheme.
  
  It is interesting that there is one special value $\gamma'=k$, i.e.~$\lambda_*=1/3$ or $\kappa=0$, for which we can find a closed form expression for $f$:
\EQ{
f=\frac{h^{2/k}-\Lambda^{2/k}}{h^{2/k}+\Lambda^{2/k}}
}
and so
\EQ{
\delta{\cal E}(h)=-\frac{h^2k}{2\pi}\cdot\frac{h^{2/k}-\Lambda^{2/k}}{h^{2/k}+\Lambda^{2/k}}\ .
}
The reason that the result has an extra simplicity here is interesting. The $\SU(2)_k$ WZW model can be expressed in terms of a compact boson of radius $R_\star = \frac{1}{\sqrt{2 k}}$ together with $\mathbb{Z}_k$ parafermions.  The interpretation of the UV parameter $\lambda_\star$ is to change the radius of the scalar according to $\lambda_\star = \frac{R_\star^2- R^2 }{R_\star^2 + R^2} $.   The special value $\gamma'=k$ corresponds to the radius $R = \frac{1}{\sqrt{2} } R_\star$ and this corresponds to the free fermion point in the moduli space of $c=1$ theories \cite{Ginsparg:1987eb}.

\subsection{The finite density TBA}

We now describe how the ground state energy can be calculated from the S-matrix. One can describe the coupling of the system at the S-matrix level by a TBA system. Since the S-matrix is non-diagonal this is a complicated set of coupled integral equations with a set of ``magnon" terms. We write down this coupled system in appendix \ref{a3}.
However, when the chemical potential is switched on, the ground state fills up with particles that carry positive $\U(1)$ charge. The S-matrix takes the form \eqref{rrp} and, with some choice of basis, the particles with positive $\U(1)$ charge are solitons of the sine-Gordon part. The S-matrix element of these states is the product of \eqref{g4r} and \eqref{sro}. These are still some complications from the magnons of the RSOS part described in appendix \ref{a3} which effectively lead to the shift $k\to k-2$. So the effective S-matrix element is precisely the hole-hole $S$-matrix element of the spin chain \eqref{sh1}.

Since only one of particle condenses in the vacuum, the resulting state can be described in a rather simple way in terms of the energy of particle states of a given rapidity $\epsilon(\theta)$ and the scattering phase of the particles extracted from the S-matrix element for the scattering of two particles
\EQ{
R(\theta)=\delta(\theta)-i\frac d{d\theta}\log S_{2s,2s}^h(\theta)\ ,
\label{cje}
}
where $S_{2s,2s}^h(\theta)$ is the hole-hole S-matrix \eqref{sh1}. The TBA equation equation takes the form
\EQ{
h-M\cosh\theta=\int_{-B}^Bd\theta'\,R(\theta-\theta')\epsilon(\theta')\ ,
\label{gew}
}
where the ``Fermi energy" is determined by the boundary conditions
\EQ{
\epsilon(\pm B)=0\ .
}

The solution of \eqref{gew} determines the ground state energy density shift via
\EQ{
{\cal E}(h)-{\cal E}(0)=-\frac M{2\pi}\int_{-B}^Bd\theta\,\cosh\theta\epsilon(\theta)\ .
}

We will present the expansion of $\delta{\cal E}(h)$ in the UV $h\gg M$ as a recipe and refer to \cite{Zamolodchikov:1995xk} for the details. In the following we will need the Fourier transform of $R(\theta)$ in \eqref{cje},
\EQ{
\tilde R(\omega)=\int_{-\infty}^\infty d\theta\,e^{i\omega\theta}R(\theta)=\frac{\sinh[\pi\omega(\gamma'+k)/2]\tanh[\pi\omega/2]}{2\sinh[\pi\omega\gamma'/2]\sinh[\pi\omega k/2]}
\ ,
\label{ssa}
}
and also its decomposition 
\EQ{
\tilde R(\omega)=\frac1{G_+(\omega)G_-(\omega)}\ ,
}
where $G_\pm(\omega)$ are analytic in the upper/lower half planes, respectively. In particular, let us define the quantity
\EQ{
\rho(\omega)=\frac{1-i\omega}{1+i\omega}\cdot\frac{G_-(\omega)}{G_+(\omega)}\ ,
\label{drh}
}
which has poles up the imaginary axis.

The expression for the ground state energy takes the form
\EQ{
{\cal E}(h)-{\cal E}(0)=-\frac{h^2}{2\pi}\cdot\frac{u(i)}{\tilde R(0)}\Big[1-\int_{{\cal C}_+}\frac{d\omega}{2\pi i}\,\frac{e^{2i\omega B}}{\omega-i}\cdot\rho(\omega)u(\omega)\Big]\ ,
\label{zap}
}
where the contour encircles all the poles on the positive imaginary axis. The key function $\U(\omega)$ satisfies the integral equation
\EQ{
u(\omega)=\frac i\omega+\int_{{\cal C}_+}\frac{d\omega'}{2\pi i}\,\frac{e^{2i\omega' B}}{\omega'+\omega}\cdot\rho(\omega')u(\omega')\ ,
\label{ieq}
}
subject to the condition
\EQ{
u(i)=\frac {Me^B}{2h}\cdot\frac{G_+(i)}{G_+(0)}\ ,
\label{bcx}
}
which determines the Fermi energy $B$.

The pole at $\omega=i$ in \eqref{zap} gives ${\cal E}(0)$ on the left-hand side
\EQ{
{\cal E}(0)=-\frac{h^2}{2\pi}\cdot\frac{u(i)}{\tilde R(0)}\,e^{-2B}\rho(i)u(i)=\frac{M^2}{4\pi}G_+(i)^2\rho(i)\ .
} 

The other poles of $\rho(\omega)$ up the imaginary axis give the $h$-dependent contribution on the left-hand side. If the poles are at $\omega=i\omega_n$ then 
\EQ{
\delta{\cal E}(h)=-\frac{h^2}{2\pi}\cdot \frac1{\tilde R(0)}\Big(1+\sum_n\frac{r_n}{1+\omega_n}\Big)\Big(1+\sum_n\frac{r_n}{1-\omega_n}\Big)\ ,
\label{ne2}
}
where we have defined the quantities
\EQ{
r_n=ie^{-2\omega_nB}u(i\omega_n)\text{Res}_{i\omega_n}\rho(\omega)\ .
}
The integral equation \eqref{ieq}, become a set of algebraic equations for the $r_n$:
\EQ{
r_n=-e^{-2\omega_nB}a_n\Big(1+\sum_m\frac{\omega_nr_m}{\omega_n+\omega_m}\Big)\ ,
\label{ne3}
}
and, finally, the condition \eqref{bcx} becomes
\EQ{
1+\sum_n\frac{r_n}{1+\omega_n}=\frac{Me^B}{2h}\frac{G_+(i)}{G_+(0)}\ .
\label{ne1}
}

From the kernel \eqref{ssa}, we have
\EQ{
G_-(\omega)=\sqrt{\frac{2\pi(\gamma'+k)}{\gamma' k}}\frac{\Gamma(i\omega(\gamma'+k)/2)\Gamma(i\omega/2)\exp(ib\omega/2)}{\Gamma(i\gamma'\omega/2)\Gamma(i\omega k/2)\Gamma(1/2+i\omega/2)}\ ,
}
with  $G_+(\omega)=G_-(-\omega)$ and where $b=\gamma'\log\gamma'+k\log k-(\gamma'+k)\log(\gamma'+k)$ is fixed by demanding that $G_+(\omega)\to 1$ as $\omega \to \infty$ everywhere away from the imaginary axis. It now apparent that $\rho(\omega)$ in \eqref{drh} has poles  due to the zeros of $G_+(\omega)$ on the imaginary axis at $2ni/(\gamma'+k)=in\delta$ and $2i(n+1)$, $n=1,2,\ldots$.

Now we can consider the scaling limit \eqref{scl} where various simplifications occur. In this limit $k\to\infty$ and $\gamma'/k$ is fixed and so only the poles at $\omega=in\delta$ are relevant.In the scaling limit since the surviving $\omega_n$ scale as $k^{-1}$, in \eqref{ne2} we can neglect the $\omega_n$'s in the denominators:
\EQ{
\delta{\cal E}(h)=-\frac{h^2\gamma' k}{\pi(\gamma'+k)}\cdot f\ ,
\label{ne5}
}
as in \eqref{rff}, with
\EQ{
f=\Big(1+\sum_{n=1}^\infty r_n\Big)^2\ .
}
In the above, we have used the fact that
\EQ{
\tilde R(0)=\frac{\gamma'+k}{2\gamma' k}\ .
}

The algebraic equations \eqref{ne3} become
\EQ{
r_n=-Q^{2n}a_n\Big(1+\sum_{m=1}^\infty \frac{nr_m}{n+m}\Big)\ ,
\label{ne6}
}
where we have defined 
\EQ{
Q=e^{-\delta B}\ .
}
The condition \eqref{ne1} becomes
\EQ{ 
1+ \sum_n r_n   =   \frac{M e^B\sqrt{2 \pi \delta }}{4 h\sqrt{1- \kappa^2}}
}
and since we are in the UV where $B$ is very large we have that
\EQ{
Q=\Big(\frac{\sqrt{\pi\delta}}{2^{3/2}\sqrt{1-\kappa^2}}\cdot\frac Mh\Big)^\delta\ .
\label{ne7}
}
It is apparent that the $r_n$ are infinite polynomials in $Q$ with lowest power $2n$. The quantities $a_n$ are found to be 
\EQ{
a_n=\Big(\frac{\kappa^2-1}4\Big)^{1-n} \Big(\frac{1-\kappa}{1+\kappa}\Big)^{n\kappa}\prod_{m=1}^{n-1}\frac{(2m-n)^2-\kappa^2n^2}{4m^2}\ ,
}
in the scaling limit.

It is remarkable that the the series $f(Q)$ defined by the recipe above agrees exactly with the series defined by perturbation theory in \eqref{rff}. This equivalence is highly non-trivial and as a by-product determines the mass gap relation between the lambda parameter and mass $M$ in the scaling limit:
\EQ{
\Lambda=\frac14\sqrt{\frac{\pi(\gamma'+k)}{\gamma' k}}\cdot M\ .
}

\section{Discussion}

Our main goal was to test the idea that anisotropic and Yang-Baxter deformed lambda and sigma models can be regularized in a way that preserves integrability as a quantum spin chain in the  light-cone lattice formalism. The hope is then that the spin chain admits a continuum limit that then defines the quantum lambda or sigma model precisely. This would provide a way calculate correlation functions, for instance. What we found was a subtle picture: anisotropic lambda models with $\lambda\geq\xi$, do indeed admit a spin chain formulation in the form of a higher spin XXZ chain in the paramagnetic regime. In this regime, the spin chain is gapless and a continuum limit exists. We showed that the S-matrix and spectrum agrees with what one expected on the basis of symmetries. In the other regime, $\lambda<\xi$, the picture was subtly different. The expected spin chain  was again the higher spin XXZ chain but now in the anti-ferromagnetic regime. This regime has a gap and so no continuum limit exists. In addition, the S-matrix of the excitations (which we wrote down for the first time in the higher spin chain) was not precisely what one expected of the continuum theory. The RSOS part did not have the right sigma model limit. So either the sigma model limit itself is more subtle in this regime, or the lambda model is not consistently regularized as a spin chain. However, the situation in the $\lambda<\xi$ regime lends weight to the idea that theories with a cyclic RG flow do not admit a continuum limit  \cite{LeClair:2004ps}.

The results we have obtained for the lambda model have implications also for sigma models. They suggest that the anisotropic $\SU(2)$ PCM in \eqref{dpcm} with deformation $(\beta,\beta,\alpha)$ define continuum, integrable QFTs when $\beta<\alpha$, The S-matrix takes the form of \eqref{sm8} in the limit $k\to\infty$ and an IRF-to-vertex transformation on the RSOS part:
\be
S = S_\text{SG}(\theta; \gamma' ) \otimes S_{\SU(2)}(\theta)\ ,
\label{sm18}
\ee
where the second block is the $\SU(2)$ invariant block and
\EQ{
\gamma'=\frac1{\sqrt{4\alpha(\alpha-\beta)}}\ ,
}
and RG invariant of the sigma model. This is a model that lies in the class studied by Fateev \cite{Fateev:1996ea}. On the other hand, in the $\beta>\alpha$ regime, the sigma model has a cyclic RG and, therefore, probably does not admit a continuum limit at the quantum level. The same conclusion befalls the Yang-Baxter sigma model to which it is equivalent in a periodic space.

When one goes beyond $\SU(2)$, the only kinds of integrable deformations of the PCM are of the Yang-Baxter type: the anisotropic deformations do not generalize. They also have cyclic RG flows and therefore it seems likely that they do not define continuum, integrable QFTs.

\vspace{2cm}

\section*{Acknowledgements}

\noindent
CA and DP are supported by STFC studentships. TJH is supported by STFC grant ST/P00055X/1.
DT is supported by a Royal Society University Research Fellowship (URF150185)  and in part by STFC grant ST/P00055X/1.

\appendix
\appendixpage

\section{TBA systems}\label{a3}

In this section, we summarize the full TBA equations associated to an RSOS S-matrix block with the goal of explaining the shift $k\to k-2$ that is crucial to our analysis. We will consider the TBA system for the UV safe regime, where the physical S matrix takes the form \eqref{rrp}. A full discussion of the TBA equations and an explanation of the shift we are after is in \cite{Hollowood:1993ac}.

The TBA equations describe the wave functions for configurations of $N$ particles of the theory in a periodic finite region $x\sim x+L$ in the thermodynamic limit $N\to\infty$, $L\to\infty$.
Since the shift in $k$ occurs in the RSOS piece, we will simplify the analysis and just write TBA equations where the scattering in the sine-Gordon block is for identical particles (in other words we will avoid introducing the magnons in this factor). The TBA equations are written in terms of the density of particles $\sigma(\theta)$ and their holes $\tilde\sigma(\theta)$. The fact that the scattering is non-diagonal in the RSOS factor means that their are a series of magnon strings of length $p=1,2,\ldots,k$ with their own desnities of occupied and un-occupied states, $r_p(\theta)$ and $\tilde r_p(\theta)$.
The TBA equations are then the coupled equations for the densities
\EQ{
&\frac m{2\pi}+\Gamma^{(k+2,\gamma')}\ast\sigma(\theta)-\sum_{p=1}^ka_p^{(k+2)}\ast r_p(\theta)=\sigma(\theta)+\tilde\sigma(\theta)\ ,\\
&\tilde r_p(\theta)+\sum_{q=1}^kA_{pq}^{(k+2)}\ast r_q(\theta)=a_p^{(k+2)}\ast\sigma(\theta)\ .
\label{ftba}
}
The equations are written using the convolution $f\ast g(\theta)=\int_{-\infty}^\infty d\theta'\,f(\theta-\theta')g(\theta')$ and the various kernels are
\EQ{
\Gamma^{(k+2,\gamma')}(\theta)&=\frac1{2\pi i}\frac d{d\theta}\log(S(\theta))\\ &=\delta(\theta)-\frac1\pi\int_{-\infty}^\infty\frac{dx}{2\pi}\,e^{2i\theta x/\pi}\,\frac{\tanh x\sinh[(k+2+\gamma')x]}
{\sinh[(k+2)x]\sinh[\gamma' x]}\ ,\\
a_p(\theta)&=\frac2\pi\int_{-\infty}^\infty\frac{dx}{2\pi}\,e^{2i\theta x/\pi}\,\frac{\sinh[(k+2-p)x]}{\sinh[(k+2)x]}\ ,\\
A_{pq}^{(k+2)}(\theta)&=\frac2\pi\int_{-\infty}^\infty\frac{dx}{2\pi}\,e^{2i\theta x/\pi}\,\frac{2\sinh(px)\sinh[(k+2-q)x]\coth x}{\sinh[(k+2)x]}\ .
}
It is simple to show that the kernel $\Gamma^{(k,\gamma')}(\theta)$, i.e.~without the shift in $k$, is precisely the hole-hole S-matrix in \eqref{sh1} once we relate the Fourier coordinates by $x=\gamma\omega/2$. If we set all the magnon densities $r_p=0$, then we have a mismatch $k$ versus $k+2$ between the TBA equation of the spin chain and the TBA equation of the physical scattering matrix. This is what we need to explain.

The TBA wave function describes a state with spin that can be related to the zero modes of the density of particles and magnons:
\EQ{
j=\frac L2\Big(\hat\sigma(0)-2\sum_{p=1}^k p\hat r_p(0)\Big)\ ,
}
where $\hat f(x)$ are the Fourier transforms:
\EQ{
f(\theta)=\frac 2\pi\int_{-\infty}^\infty\frac{dx}{2\pi}\,e^{2i\theta x/\pi}\hat f(x)\ .
}
Using the Fourier transform of the magnon equation \eqref{ftba} for $p=k$, we find
\EQ{
j=L\frac{k+2}4\hat{\tilde r}_k(0)\ .
}
In the RSOS sector, the allowed spins are bounded by $j\leq k/2$. The implication is that in the thermodynamic limit, $L\to\infty$, the number of $k$ string holes must go to zero: $\tilde r_k(\theta)\to0$. 
This allows us to eliminate $r_k(\theta)$ from the TBA equations using
\EQ{
r_k(\theta)=-\sum_{p=1}^{k-1}\big[A_{kk}^{(k+2)}\big]^{-1}\ast A_{kp}^{(k+2)}\ast r_p(\theta)+\big[A_{kk}^{(k+2)}\big]^{-1}\ast a_{k}^{(k+2)}\ast\sigma(\theta)\ .
}
where 
\EQ{
[f]^{-1}(\theta)=\frac 2\pi\int_{-\infty}^\infty\frac{dx}{2\pi}\,e^{2i\theta x/\pi}\frac1{\hat f(x)}\ .
}
What results are a set of effective TBA equations valid in the thermodynamic limit:
\EQ{
&\frac m{2\pi}+\Gamma^{(k,\gamma')}\ast\sigma(\theta)-\sum_{p=1}^{k-1}a_p^{(k)}\ast r_p(\theta)=\sigma(\theta)+\tilde\sigma(\theta)\ ,\\
&\tilde r_p(\theta)+\sum_{q=1}^{k-1}A_{pq}^{(k)}\ast r_q(\theta)=a_p^{(k)}\ast\sigma(\theta)\ .
}
So effectively what has happened relative to \eqref{ftba} is that the string label on the magnons now only goes $1,2,\ldots,k-1$ and the kernels have the shift $k\to k-2$. In particular the scattering of the highest weight states is now governed by an effective S-matrix element that matches the spin chain hole-hole element in \eqref{sh1} exactly.

\section{S-matrix bootstrap}\label{a2} 
\pgfdeclarelayer{background layer} 
\pgfdeclarelayer{foreground layer} 
\pgfsetlayers{background layer,main,foreground layer}

In this section, we explain more explicitly the S-matrix bootstrap that we use in section \ref{s6}.

The most non-trivial aspect of S-matrix theory is the analytic structure and its explanation in terms of bound states and anomalous thresholds. Let us suppose that the states have masses $m_a$ are come in multiplets described by vector spaces $V_a$. Bound states are signalled by simple poles of the S-matrix on the physical strip, the region $0<\IM\,\theta<\pi$. For integrable field theories these poles occur at purely imaginary values: if a bound state corresponding to 
particle $V_c$ is exchanged
in the direct channel then $S_{ab}(\theta)$ has a simple pole at the imaginary value $\theta=iu_{ab}^c$, with $0<u_{ab}^c<\pi$ where
\EQ{
m_c^2=m_a^2+m_b^2+2m_am_b\cos u_{ab}^c\ .
}
Note that if $c$ is a bound state of $a$ and $b$, then $a$ is a bound state of $c$ and $\bar b$, the anti-particle of $b$, and $b$ is a bound state of $c$ and $\bar a$,
\EQ{
u_{ab}^c+u_{b\bar c}^{\bar a}+u_{a\bar c}^{\bar b}=2\pi\ ,
}
as illustrated in figure~\ref{f3} 
\FIG{
\begin{tikzpicture} [line width=1.5pt,inner sep=2mm,
place/.style={circle,draw=blue!50,fill=blue!20,thick},proj/.style={circle,draw=red!50,fill=red!20,thick}]
\node at (2,2) [proj] (p1) {};
\node at (0.4,0.4) (i1) {$a$};
\node at (3.6,0.4) (i2)  {$b$};
\node at (2,4) (i3) {$c$};
\draw[->]  (i1)  --  (p1);
\draw[->]  (i2)  --  (p1);
\draw[<-]  (i3)  --  (p1);
\node at (2,0.3) {$u_{ab}^c$};
\node at (0.8,2.5) {$u_{a\bar c}^{\bar b}$};
\node at (3.2,2.5) {$u_{b\bar c}^{\bar a}$};
\end{tikzpicture}
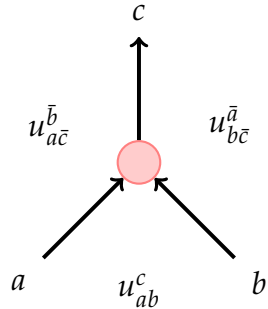
\caption{\small The rapidity angles for the 3-point functions.}
\label{f3}
}

At the bound-state pole $\theta_{12}=iu_{ab}^c$ the bound state $V_c$ must lie in the product of representations of the incoming states
\EQ{
V_a\otimes V_b=V_c\oplus V_c^\perp\ .
}
The complement $V_c^\perp$ lies in the kernel of $\text{Res}\,S_{ab}(iu_{ab}^c)$:
\EQ{
\text{Res}\,S_{ab}(iu_{ab}^c):\; V_c^\perp\longrightarrow0\ .
\label{ipm}
}
In general, the representation $V_c$ is reducible under the underlying symmetry algebra and let us we write the decomposition as 
\EQ{
V_c=\oplus_jV_c^{(j)}\ ,
}
then near the pole we have
\EQ{
S_{ab}(\theta)\thicksim \frac i{\theta-iu_{ab}^c}R_{ab}\ ,\qquad R_{ab}=\sum_jr_j\,\Pr_j\ ,
\label{fgf}
}
where $\Pr_j$ is a projector onto $V_c^{(j)}\subset V_a\otimes V_b$.  
In \eqref{fgf}, the numbers $r_j$ are required to be real, and unitarity of the underlying QFT dictates the sign. In simple cases, the sign is related to the parity of the bound state as found by Karowski~\cite{Karowski:1978ps}. The coupling of asymptotic states to the bound state is illustrated in figure~\ref{f4}. 
 
\FIG{
\begin{tikzpicture} [line width=1.5pt,inner sep=2mm,
place/.style={circle,draw=blue!50,fill=blue!20,thick},proj/.style={circle,draw=red!50,fill=red!20,thick}]
\begin{pgfonlayer}{foreground layer}
\node at (1,1) [proj] (p1) {};
\node at (1,3) [proj] (p2) {};
\node at (-0.5,-0.5) (i1) {$a$};
\node at (2.5,-0.5) (i2) {$b$};
\node at (-0.5,4.5) (i3) {$b$};
\node at (2.5,4.5) (i4) {$a$};
\node at (0.5,2) {$c$};
\node at (3,2.8) (s1) {$R_{ab}^{1/2}$};
\draw[->,thin] (s1)  --  (p2);
\node at (3,1.2) (s3) {$R_{ab}^{1/2}$};
\draw[->,thin] (s3)  --  (p1);
\end{pgfonlayer}
\draw[-] (i1)  --  (p1);
\draw[-] (i2)  --  (p1);
\draw[<-] (i3)  --  (p2);
\draw[<-] (i4)  --  (p2);
\draw[-,red] (p2)  --  (p1);
\end{tikzpicture}
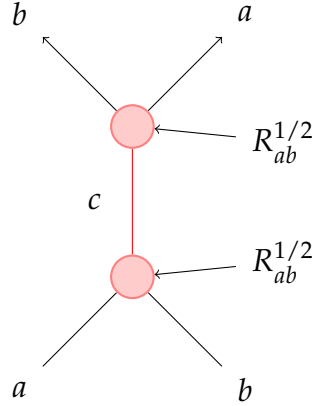
\caption{\small The anatomy of the S-matrix in the vicinity of a bound state pole.}
\label{f4}
}

The fact that $c$ can appear as a bound state of $a$ and $b$ means that the S-matrix elements of $c$ with other states, say $d$, can be written in terms of those of $a$ and $b$. This is the essence of the bootstrap, or fusion, programme. The relation between the S-matrix elements can be written concretely as
\EQ{
S_{dc}(\theta)=R_{ab}^{1/2}\,S_{db}(\theta+i\bar u_{b\bar c}^{\bar a})S_{da}(\theta-i\bar u_{a\bar c}^{\bar b})\,R_{ab}^{-1/2}\ ,
\label{bfs}
}
where $\bar u_{ab}^c=\pi-u_{ab}^c$. This expression follows in an obvious way from the equality illustrated in figure \ref{f5}. 
\FIG{
\begin{tikzpicture} [scale=0.7,line width=1.5pt,inner sep=2mm,
place/.style={circle,draw=blue!50,fill=blue!20,thick},proj/.style={circle,draw=red!50,fill=red!20,thick}]
\begin{pgfonlayer}{foreground layer}
\node at (1.5,1.5) [place] (sm1) {}; 
\node at (1.8,3.4) [proj] (sm2) {}; 
\node at (3.5,2.5) [place] (sm3) {}; 
\node at (10.7,3.4) [place] (sm10) {}; 
\node at (11.5,1.7) [proj] (sm11) {};
\node at (6.5,2.5) {$=$};
\end{pgfonlayer}
\node at (11,0) (k1) {$a$};
\draw[-] (k1)  --  (sm11);
\node at (13,0.7) (k2) {$b$};
\draw[-]  (k2)  --  (sm11);
\node at (8.7,2.3) (k8) {$d$};
\draw[->] (k8)  --  (12.8,4.5);
\node at (9.8,5.5) (k3) {$c$};
\draw[->,red] (sm11)  --  (k3);
\node at (0.8,5.5) (k4) {$c$};
\draw[->,red] (sm2)  --  (k4);
\node at (5,1.7) (k5) {$b$};
\draw[-] (k5)  --  (sm2);
\node at (1.25,0) (k6) {$a$};
\node at (0,0.8) (k7) {$d$};
\draw[->] (k7)  --   (4.7,3.15);
\draw[-] (k6)  --  (sm2);
\node at (-1,2) (t1) {$\theta-i\bar u_{a\bar c}^{\bar b}$};
\draw[->,thin] (t1)  --  (sm1);
\node at (3.5,4.3) (t2) {$\theta+i\bar u_{b\bar c}^{\bar a}$};
\draw[->,thin] (t2)  --  (sm3);
\node at (10.2,2) (t3) {$\theta$};
\draw[->,thin] (t3)  --  (sm10);
\node at (-1,3.5) (q1) {$R_{ab}^{1/2}$};
\node at (14.4,2.5) (q2) {$R_{ab}^{1/2}$};
\draw[->,thin] (q1)  --  (sm2);
\draw[->,thin] (q2)  --  (sm11);
\end{tikzpicture}
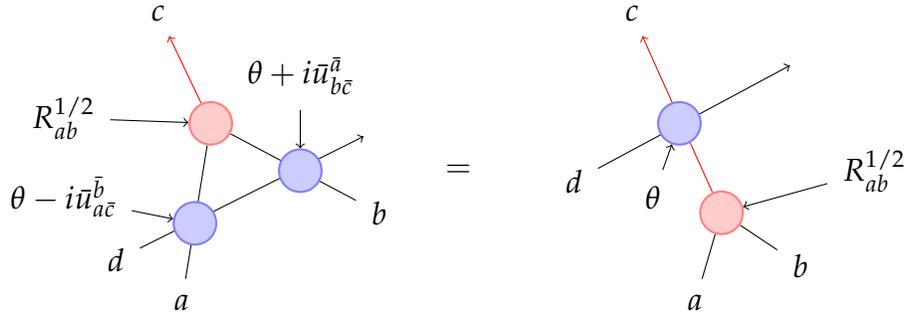
\caption{The bootstrap equations result from the equality of the diagrams above. One understands these diagrams in terms of localized wavepackets. The higher spin conserved charges implied by integrability can be used to move the trajectory of particle $d$ so that it either interacts with bound state $c$ or the particles $a$ and $b$ of which $c$ is composed. In order to isolate $S_{dc}(\theta)$ one has to act on the right by $R_{ab}^{-1/2}$.}
\label{f5}
}

The Yang-Baxter equation near the pole $\theta=iu_{ab}^c$, takes the form
\EQ{
R_{ab}S_{db}(\theta+i\bar u_{b\bar c}^{\bar a})S_{da}(\theta-i\bar u_{a\bar c}^{\bar b})=S_{da}(\theta-i\bar u_{a\bar c}^{\bar b})S_{db}(\theta+i\bar u_{b\bar c}^{\bar a})R_{ab}\ ,
}
can used to re-write the expression \eqref{bfs} as
\EQ{
S_{dc}(\theta)=R_{ab}^{-1/2}\,S_{da}(\theta-i\bar u_{a\bar c}^{\bar b})S_{db}(\theta+i\bar u_{b\bar c}^{\bar a})\,R_{ab}^{1/2}\ .
\label{bfs2}
}

\end{document}